\documentclass[aps,twocolumn,prd,showpacs,nofootinbib]{revtex4}
\usepackage{amsmath}
\usepackage{graphicx}
\usepackage{dcolumn}
\usepackage{bm}
\usepackage{amssymb}
\usepackage{latexsym}

\def\setC{\mathbb{C}}

\def\setR{\mathbb{R}}

\newcommand{\dd}{\mathrm{d}}
\newcommand{\ee}{\mathrm{e}}

\newcommand{\ie}{\textsl{i.e.~}}

\newcommand{\eg}{\textsl{e.g.~}}
\newcommand{\etal}{\textsl{et al.~}}
\newcommand{\apriori}{\textsl{a priori~}}

\newcommand{\Ups}{\Upsilon}

\newcommand{\mP}{m_{_\mathrm{Pl}}}
\newcommand{\GReCO}{${\cal G}\setR\varepsilon\setC{\cal O}$}

\def\spose#1{\hbox to 0pt{#1\hss}}
\def\lta{\mathrel{\spose{\lower 3pt\hbox{$\mathchar"218$}}
     \raise 2.0pt\hbox{$\mathchar"13C$}}}
\def\gta{\mathrel{\spose{\lower 3pt\hbox{$\mathchar"218$}}
     \raise 2.0pt\hbox{$\mathchar"13E$}}}

\newcommand{\ellC}{\ell_{_{\mathrm C}}}
\newcommand{\ellH}{\ell_{_{\mathrm H}}}
\newcommand{\cS}{c_{_{\mathrm S}}}
\newcommand{\cs}{c_{_{\mathrm S}}}

\newcommand{\Hu}{{\cal H}} \newcommand{\Ka}{{\cal K}}

\begin{document}

\title{Parametric amplification of metric fluctuations through a
bouncing phase}

\author{J\'er\^ome Martin} 
\email{jmartin@iap.fr}
\affiliation{Institut d'Astrophysique de
Paris, \GReCO, FRE 2435-CNRS, 98bis boulevard Arago, 75014 Paris,
France}

\author{Patrick Peter}
\email{peter@iap.fr}
\affiliation{Institut d'Astrophysique de
Paris, \GReCO, FRE 2435-CNRS, 98bis boulevard Arago, 75014 Paris,
France}

\date{October 6$^\mathrm{th}$, 2003}

\begin{abstract}
We clarify the properties of the behavior of classical cosmological
perturbations when the Universe experiences a bounce. This is done in
the simplest possible case for which gravity is described by general
relativity and the matter content has a single component, namely a
scalar field in a closed geometry. We show in particular that the
spectrum of scalar perturbations can be affected by the bounce in a
way that may depend on the wave number, even in the large scale
limit. This may have important implications for string motivated
models of the early Universe.
\end{abstract}

\maketitle

\section{Introduction}

Although the idea that the universe could have experienced a bounce in
its remote past is old~\cite{tolman,seventies}, it has recently come
under new scrutiny~\cite{bounce,ppnpn1,GT,GGV} with the advent of
string motivated scenarios of the pre big bang
kind~\cite{PBB,ekp}. The main reason for this renewed interest is the
fact that the most popular extensions of the standard model of high
energy physics, such as string or M theory, when applied to cosmology,
\ie in a four dimensional time dependent background, can lead to
solutions with bouncing scale factors (see, \eg
Refs.~\cite{bounce,GGV,PBB} and references therein).

A crucial property to decide whether these new models can be turned
into realistic alternatives to the inflationary paradigm which, so
far, has been so successful is the behavior of cosmological
perturbations around these bouncing backgrounds. In particular, an
important test is to calculate the evolution of the power spectrum of
primordial fluctuations through the bounce in order to see whether it
can be made close to scale invariance, \ie if there is any
possibility, given the prebounce era, to get a Harrison-Zel'dovich
power spectrum.

{}From a technical point of view, the previous question is a
nontrivial problem. Simple models, based on general relativity with
flat spatial sections~\cite{ekp}, lead to the existence of a curvature
singularity at the bounce itself and therefore do not seem to
represent viable physical models. In addition, it is difficult to
understand how meaningful a perturbative scheme around a singular
solution would be (see however Ref.~\cite{5D}), so we shall assume
that the question of the calculation of the cosmological perturbations
cannot be addressed in this way (see Ref.~\cite{noekp} for more
detailed discussions). In fact, it is believed that, in the vicinity
of the bounce, string corrections become
important~\cite{PBB}. Typically, these corrections add to the gravity
sector terms such as $R^2$, $R_{\mu \nu \rho \lambda }R^{\mu \nu \rho
\lambda }$, where $R_{\mu \nu \rho \lambda }$ is the Riemann tensor
and $R$ the curvature scalar~\cite{string}, and higher order terms in
the curvature. The effect of these terms is, except in some specific
instances~\cite{5D}, to smooth out the singularity~\cite{regcos} (see
also Ref.~\cite{PBB}); this is, from a physical point of view,
satisfactory and expected. Unfortunately, one can show that the
process of adding up more and more high curvature corrections does not
lead to convergence towards a single solution, \ie this solution
explicitly depends on the choice of the stringy
corrections~\cite{addR}. This means that, at each order, the bouncing
scale factor looks completely different from the scale factor obtained
at the previous order. Nevertheless, at a fixed order in the string
corrections, one can in principle compute how the perturbations
propagate through the bounce. The main disadvantage of the procedure
is that it renders the computation extremely complicated and only
numerical calculations are available to make the problem tractable.

An important point to be noticed is that, as already mentioned above,
most models assume that the spatial sections are flat all the time
whereas, at the bounce, the curvature term is expected to play a
crucial role. Therefore, it seems that for a bouncing universe, one
cannot just throw away the curvature term because it does not play a
significant role as it is the case for an inflationary universe. In
fact, in that regard, the situation is the opposite of inflation:
during the final stages of inflation, one can safely assume flat
spatial sections because the three-curvature is getting more and more
negligible as time passes, whereas even though the curvature may be
negligible either in the remote past or in the future of the bounce,
it has almost certainly no reason to be so in general.

A way out to the previous difficulties, which would permit to
undertake a tractable analytical calculation of the power spectrum, is
the following. Far from the bounce, one usually considers the
situation for which the curvature is small, even though the
implementation of this particular point may not be in itself a trivial
task. In this case, one can consider that, during the contracting and
expanding phases, the spatial sections are essentially flat so that
the well known results stemming from the theory of cosmological
perturbations can be straightforwardly applied. Then, the main
question becomes the effect of the bounce itself on the pre-bounce
power spectrum. Technically, this problem can be formulated as
follows~\cite{DV}. Before the bounce, the perturbations are
characterized by two modes, a dominant (denoted by $D$) and a
sub-dominant ($S$) one. With $k$ representing the comoving wave number
of a given Fourier mode, we write the $k$ dependence of these modes as
$D_-(k)$ and $S_-(k)$. In the same manner, after the bounce, one
decomposes the perturbation as $D_+(k)$ and $S_+(k)$. The effect of
the bounce is then entirely encoded into the form of the transition
matrix $T(k)$ defined by
\begin{equation}
\begin{pmatrix}
D_+\cr S_+
\end{pmatrix}
=
\begin{pmatrix} 
T_{11} & T_{12} \cr T_{21} & T_{22}
\end{pmatrix}
\begin{pmatrix}
D_- \cr S_-
\end{pmatrix}\, .
\label{transT}
\end{equation}
The mode of interest is of course the dominant mode in the expanding
phase, $D_+(k)=T_{11}(k)D_-(k)+T_{12}(k)S_-(k)$. This equation
corresponds to a general ``$k$-mode mixing,'' \ie the dominant mode
after the bounce is a general linear combination of the dominant and
sub-dominant modes before the bounce.\footnote{We have introduced the
notation ``$k$-mode mixing'' in order to emphasize that this mode
mixing is valid for a fixed Fourier mode and should not be confused
with the mode mixing coming from nonlinearities which involves modes
with different wave numbers.}

{\it A priori}, various different situations can occur: the dominant
mode in the contracting phase could acquire a scale invariant spectrum
which is not conveyed to the dominant mode in the expanding phase
because it turns out that $T_{11}=0$ and $T_{12}\neq 0$ (``$k$-mode
inversion,'' the scale invariant piece is passed to the ``wrong mode''
in the expanding phase); this is for instance what occurs if one
applies the usual Israel junction conditions, known to apply for other
cosmological transitions, at the bounce point~\cite{BF}.

Another possibility is that the dominant mode in the contracting phase
be scale invariant but that this property is lost through the bounce
due to a nontrivial $k$ dependence of the coefficient $T_{11}$. Note
that the opposite situation may also occur, for which the spectrum is
initially not scale invariant but is turned into it because of a
nontrivial $k$ dependence of the transition matrix. In fact, the
common view concerning these last possibilities is that, for scales of
astrophysical interest today, the bounce, lasting a short time, is
expected to have no noticeable effect on those large
scales~\cite{PBB}. This is sometimes argued to come from general
arguments such as ``causality,'' a point which is discussed thoroughly
in Ref.~\cite{MP}. For instance, this is a basic assumption in the
perturbation spectrum calculations in the pre big bang
scenario~\cite{PBB}. Technically, this means that the transfer matrix
is assumed not to depend on $k$~\cite{DV}. Within this framework, the
goal reduces to finding situations for which a scale invariant
spectrum is produced in the contracting phase, and to ensure that this
spectrum is passed to the dominant mode in the expanding phase, \ie to
insure that the matching conditions at the bounce do not imply a
$k$-mode inversion.

The present article aims at examining whether the assumption that the
transfer matrix is $k$-independent is generically valid or not. For
this purpose, we need to specify a class of models where bouncing
solutions are possible and which allows simple analytical treatment of
the perturbations through the bounce. We choose general relativity,
positive curvature spatial section (see the remarks above), and
describe the matter content by a scalar field; a similar strategy was
used in Ref.~\cite{GT}. We do not assume anything relative to what
happens away from the bounce, and in particular one could envisage
that there the curvature is negligible; note that, as we show below,
this implies the existence of a new phase. Therefore our closed
geometry bounce can be viewed as an example of a transition connecting
the contraction phase to the expanding phase with flat spatial
sections, as already considered in the literature.

This article is organized as follows. In the following section, we set
the precise model and derive the basic equations both for the
background and the perturbations. We then discuss how one can model a
bounce in this framework and derive an explicit form for the potential
of the scalar part of the classical perturbations (Bardeen potential)
whose properties we then examine in details. This leads us to the main
calculation of this article, namely that of the transfer matrix of
Eq.~(\ref{transT}). We show that this matrix depends on $k$ in a
nontrivial way provided that the null energy condition (NEC) is very
close to being violated at the bounce. This illustrates, by means of a
specific example, that the general argument according to which the
limited but nonvanishing bounce duration could not affect the spectrum
of long (\ie longer that the duration itself) wavelength modes, is
incorrect. We conclude by discussing this result, also showing that in
the case under consideration, the propagation of gravitational waves
(tensor modes) is qualitatively different of that of scalar modes
since the former are never affected by the bounce.

\section{Basic equations}

We assume that the background model is given by a
Friedmann-Lema\^{\i}tre-Robertson-Walker (FLRW) universe, \ie
\begin{equation}
\label{metric}
{\rm d}s^2=a^2(\eta )\biggl[-{\rm d}\eta ^2+\frac{{\rm d}r^2}{1-{\cal
K}r^2} +r^2({\rm d}\theta ^2+\sin ^2\theta {\rm d}\phi ^2)\biggr]\,
.
\end{equation}
In this equation, the constant parameter ${\cal K}$ can always be
rescaled such that ${\cal K}=0,\pm 1$ and describes the curvature of
the spatial sections. The time $\eta $ is the conformal time related
to the cosmic time by ${\rm d}t=a(\eta ){\rm d}\eta$. The matter is
described by a homogeneous scalar field $\varphi (\eta )$ and the
corresponding energy density and pressure respectively read
\begin{equation}
\rho =\frac{\varphi '^2}{2a^2}+V(\varphi )\, ,\quad 
p=\frac{\varphi '^2}{2a^2}-V(\varphi ).
\label{rhop}
\end{equation}
A prime denotes a derivative with respect to conformal time. The
function $V(\varphi )$ represents the potential of the scalar field.
Einstein equations relate the scale factor to the energy density and
pressure of the scalar field according to
\begin{eqnarray}
\frac{3}{a^2}({\cal H}^2+{\cal K}) &=& \kappa \rho \, , \label{rho} \\
-\frac{1}{a^2}(2{\cal H}'+{\cal H}^2+{\cal K})&=&\kappa p\, ,
\label{p}
\end{eqnarray}
where we have defined ${\cal H}\equiv a'/a$ and $\kappa \equiv
8\pi/m_{\rm Pl}^2$, $m_{\rm Pl}$ being the Planck mass. The quantity
$\rho +p$ is then given by
\begin{equation}
\kappa (\rho +p)=\frac{2}{a^2}{\cal H}^2\Gamma =\frac{2}{a^2}({\cal H}^2
-{\cal H}'+{\cal K})=\kappa \frac{\varphi '^2}{a^2}\, . \label{nec}
\end{equation} 
{}From the above equation, we see that the function $\Gamma (\eta )$
is defined by $\Gamma \equiv 1-{\cal H}'/{\cal H}^2 +{\cal K}/{\cal
H}^2$. It is directly related to the equation of state parameter
$\omega \equiv p/\rho $ by the following relation $\omega =(2\Gamma
/3)(1+{\cal K}/{\cal H}^2)^{-1}-1$. The function $\Gamma (\eta )$
reduces to a constant for constant equation of state and is zero in
the particular case of the de Sitter manifold. At the bounce, the
Hubble parameter vanishes, ${\cal H}=0$, while $\Hu'>0$, and therefore
the only way to preserve the null energy condition $\rho +p\geq 0$ is
to have ${\cal K}>0$. This is why, in this article, we restrict
ourselves to the case ${\cal K}>0$, \ie the spatial sections are
3-spheres. Let us also notice that, being given a bouncing scale
factor $a(\eta )$, it is sufficient to check that $\Gamma \geq 0$ at
all times in order for the scale factor to be a solution of the
Einstein equations with a single real scalar field.

A universe with closed spatial sections is characterized by two
fundamental lengths. The first length is the Hubble length defined by
$\ellH\equiv a^2/a'=a/\dot a\equiv H^{-1}$ (a dot denoting a
derivative with respect to cosmic time $t$) and the second one is the
curvature radius, $\ellC\equiv a/\sqrt{\vert {\cal K}\vert }$. The
flat limit is recovered when $\ellC\gg \ellH$ as revealed by the
equation $\vert 1-\Omega \vert =\ellH^2/ \ellC^2$, where $\Omega $ is
the ratio of the total energy density $\rho$ to the critical energy
density. When it comes to numerical applications, let us recall that
one can safely assume the preferred value~\cite{WMAP} $H_0 = 100 h
$~km$\cdot$s$^{-1}\cdot$Mpc$^{-1}$ with $h=0.71^{+0.04}_{-0.03}$,
leading to a Hubble distance scale now of $\sim 3000 h^{-1}$~Mpc~$\sim
4.2\pm 0.2$~Gpc. Moreover, with $\Omega_\mathrm{now}=1.02\pm 0.02$,
one has a curvature length, namely the scale factor as measured
now~\cite{ACK}, of order $a_0 \gta 15 h^{-1}$~Gpc (with $\Ka=1$), the
limit coming from the maximum allowed value for $\Omega_\mathrm{now}$
at one $\sigma$ level.

At the perturbed level, and in the presence of density perturbations
only, the metric takes the following form
\begin{eqnarray}
\dd s^2 & = &a^2(\eta )\Bigl\{-\left(1+2\phi \right) \dd\eta^2
+2\partial_i B\dd\eta \dd x^i \nonumber \\
& & \left. + \left[\left(1-2\psi \right) \gamma _{ij}^{(3)}
+2\nabla_i \partial_j E\right] \dd x^i \dd x^j \right\}\, , 
\label{pertmet}
\end{eqnarray}
where $\gamma _{ij}^{(3)}$ is the metric of the spatial sections
and the symbol $\nabla _i$ denotes the covariant derivative associated
with the three-dimensional metric. The eigenfunctions $f_n(x^i)$ of
the Laplace-Beltrami operator on the spatial sections satisfy 
the equation
\begin{equation}
\Delta f_n=-n(n+2)f_n\, ,
\label{laplace}
\end{equation}
where $n$ is an integer. Note at this point that it is because of our
normalization with a dimensionful scale factor $a(\eta)$, and hence
dimensionless coordinates $\left(\eta,x^i\right)$, implying a
dimensionless operator itself, that the eigenvalues of $\Delta$ are
dimensionless integer numbers; with a different convention, \ie with
$a$ dimensionless, one would have $\left[ \Delta \right] = L^{-2}$ and
an extra factor $\ellC^{-2}$ would appear in the right hand side of
Eq.~(\ref{laplace}).

The modes $n=0$, corresponding to a homogeneous deformation, and
$n=1$, being nothing but a global motion of the center of the
3-sphere, are pure gauge modes~\cite{LK}: we will accordingly consider
only values of $n$ such that $n>1$. In fact, for the relevant
cosmological parameters discussed above, one finds that the values of
$n$ corresponding to characteristic distance scales of cosmological
interest now, namely $10^{-2} h^{-1}$~Mpc~$\alt D_\mathrm{cosm} \alt
10^3 h^{-1}$~Mpc, range between $30$ and $3\times 10^6$ for the
largest possible value of the total density now. For a reasonable
value of $\Omega_\mathrm{now} \sim 1.01$, we find that $n$ is between
$60$ and $6\times 10^6$.

The scalar perturbations are described by the four functions, $\phi$,
$B$, $\psi $ and $E$ and, from them, it is possible to construct two
gauge-invariant quantities, called the Bardeen potentials, and defined
by~\cite{bardeen,mfb}
\begin{equation}
\Phi \equiv \phi +\frac{1}{a}\biggl[a(B-E')\biggr]'\, , \quad 
\Psi \equiv \psi -\frac{a'}{a}(B-E')\, .
\end{equation}
For simple form of matter with no anisotropic stress (this is the case
for a scalar field), we have $\Phi=\Psi$. Notice that the form of the
Bardeen potentials is the same whatever the curvature of the spatial
sections is. This is related to the fact that, even if ${\cal K}>0$
(or ${\cal K}<0$), the FLRW metric remains conformally flat and the
components of the perturbed Weyl tensor remain unchanged. Then,
Stewart lemma guarantees that the Bardeen potentials are still defined
by the same equations~\cite{stewart}.

For the matter sector, the scalar field is written as $\varphi +\delta
\varphi (\eta ,x^i)$ where $\delta \varphi (\eta ,x^i)$ represents the
inhomogeneous fluctuations. These fluctuations can be described by the
gauge invariant quantity $\delta \varphi ^{\rm (gi)}\equiv \delta
\varphi +\varphi '(B-E')$.

The full set of Einstein equations can be written in terms of the
gauge invariant quantities $\Phi $ and $\delta \varphi ^{\rm (gi)}$
only.  Combining these equations permits to derive a master equation
for the Bardeen potential (for $\varphi '\neq 0$) which reads
\begin{equation}
\Phi ''+2\biggl({\cal H}-\frac{\varphi ''}{\varphi '}\biggr)\Phi '
+\biggr[n(n+2)+2\biggl({\cal H}'-{\cal H}\frac{\varphi ''}{\varphi '}
-2{\cal K}\biggl)\biggr]\Phi=0\, . \label{eqBardeen}
\end{equation}
This equation can be cast into a more convenient form. For this 
purpose, one introduces a new gauge-invariant quantity, $u$, related 
to the Bardeen potential $\Phi $ by
\begin{equation}
\Phi \equiv \frac{\kappa }{2}(\rho +p)^{1/2}u =\frac{\sqrt{3\kappa }}{2}
\frac{{\cal H}}{a^2\theta }u \, ,\label{defu}
\end{equation}
where the function $\theta $ is defined by 
\begin{equation}
\label{deftheta}
\theta \equiv \frac{1}{a}\biggl(\frac{\rho }{\rho +p}\biggr)^{1/2}
\biggl(1-\frac{3{\cal K}}{\kappa \rho a^2}\biggr)^{1/2}=
\frac{1}{a}\biggl(\frac{3}{2\Gamma }\biggr)^{1/2}\, .
\end{equation}
Then, the equation of motion for the quantity $u$ takes the form
\begin{equation}
\label{eomu}
u''+\biggl[n(n+2)-\frac{\theta ''}{\theta }-3{\cal K}(1-c_{_{\rm S}}^2)
\biggr]u=0\, .
\end{equation}
In the above equation, one has
\begin{equation}
c_{_{\rm S}}^2\equiv \frac{p'}{\rho'} = -\frac{1}{3} \left( 1+
2\frac{\varphi''}{\Hu \varphi'}\right),\label{cs2}
\end{equation}
for the scalar field, when use is made of the Klein-Gordon equation
\begin{equation}
\varphi'' + 2\Hu \varphi' + a^2 \frac{\dd V(\varphi)}{\dd\varphi} = 0.
\end{equation}
The quantity $\cs$ of Eq.~(\ref{cs2}) can, in some regimes, be
interpreted as the sound velocity. Let us now see how the flat case is
recovered. The term $\theta ''/\theta $ is of order ${\cal H}^2$,
namely $\theta ''/\theta \sim a^2/\ell _{_{\rm H}}^2$. This is a
rigorous statement if the scale factor is a power law of the conformal
time, which explains the usual confusion between the potential and the
Hubble scale. However, this identification suffers from important
exceptions, particularly relevant in the present context; see the
discussion in Sec.~\ref{other}. Then the above equation can be
re-written as
\begin{equation}
u''+a^2\biggl[\frac{4\pi ^2}{\lambda ^2}-\frac{1}{\ell _{_{\rm H}}^2}-
\frac{3}{\ell _{_{\rm C}}^2}(1-c_{_{\rm S}}^2)\biggr]u\simeq 0\, , 
\end{equation}
where we have used that the physical wavelength of a mode is $\lambda
(\eta) =2\pi a(\eta ) /\sqrt{n(n+2)}$. In the limit $\ellC\gg \ellH$,
the last term of the equation becomes negligible. Then the equation of
motion for $u$ reduces to $u''+[n(n+2)-\theta ''/\theta ]u=0$ where
now $\theta $ denotes the function defined previously in
Eq.~(\ref{deftheta}) but without the term proportional to ${\cal K}$
(one can also show that, in the limit considered here, this term
becomes negligible).  Therefore, we have recovered the standard
equation, valid for $\Ka=0$.

Note that in the flat limit for which $\Ka= 0$, the mode number
$n(n+2)$ that appears in Eq.~(\ref{eomu}) appears to be a large
number, and even more so when $\Omega\to 1$, as is the case in the
usual cosmological calculations based on a period of inflation for
which the approximation $k^2 \ll 1$ is often done, permitting an
expansion in powers of $k^2$. This is not inconsistent though, because
after $N$ e-folds of inflation have happened, in a closed situation
for instance, one expects $\Omega-1\sim \ee^{-2N}$, with $N>55$ to set
the scales, and hence a gigantic value for the scale factor
normalization $a_0 = H_0^{-1} \left( \Omega -1
\right)^{-1/2}$. However, one then assumes, rightly, that the universe
is almost flat, and chooses in general a different normalization for
the scale factor, namely $a_0 = H_0^{-1}$, which is many orders of
magnitude below what it ought to be. The relevant wave numbers, seen as
eigenvalues of the Laplace-Beltrami operator, then must scale
correspondingly, \ie they are reduced by the amount
\begin{equation}
n(n+2)\to k^2 =n(n+2) \left( \Omega -1
\right),
\end{equation}
which is, indeed, much smaller than unity in any inflationary
scenario.

The equation of motion for the quantity $u$, Eq.~(\ref{eomu}), has the
traditional form of a parametric oscillator equation, \ie of a
``time-independent'' Schr\"odinger equation. However, in the case
${\cal K}\neq 0$, the effective potential cannot be written as the
second time derivative of some function ($\theta $ in the flat case)
over the same function because of the presence of the term $-3{\cal
K}(1-c_{_{\rm S}}^2)$ in the time-dependent frequency. This has for
consequence that, on ``super-horizon'' scales where the term $n(n+2)$
is negligible, the solutions are not easily found contrary to the flat
case where they are just $u=\theta $ and $u=\theta \int {\rm
d}\tau/\theta ^2$.

So far, we have discussed the quantity $u$ which is, up to some
background functions, the Bardeen potential. In the framework of
cosmological perturbations, there exists another important variable,
usually denoted $v$, that we now consider. This quantity is important
because its flat case equivalent naturally appears when one studies
cosmological perturbations of quantum-mechanical origin. In other
words, this quantity is interesting for setting up physically
well-motivated initial conditions whenever the curvature is
negligible. Its definition reads
\begin{equation}
\displaystyle
v=\frac{-a}{\sqrt{1-3{\cal K}\displaystyle\frac{1-c_{_{\rm S}}^2}{n(n+2)}}}
\left[\delta \varphi ^{\rm (gi)}+\frac{\varphi '}{{\cal H}}\Phi
-\frac{{\cal K}\varphi '}{{\cal H}^3\Gamma }\Phi \right]\, .
\label{defv}
\end{equation}
For ${\cal K}=0$, it reduces to the well-known definition. The
presence of the factor $n(n+2)$ in the definition above suggests
however that this variable, in the $\Ka\neq 0$ case, is not the
canonical field that should be quantized to get initial conditions.
This quantity involves $\delta \varphi ^{\rm (gi)}$ and $\Phi$. Since
there are related by the perturbed Einstein equations, there is in
fact only one degree of freedom as expected. The equation of motion
for $v$ reads
\begin{equation}
v''+\left[n(n+2)-\frac{z''}{z}-3{\cal K}\left(1-c_{_{\rm S}}^2\right)
\right]v=0\, ,\label{eov}
\end{equation}
where the quantity $z$ is defined by
\begin{equation}
z\equiv \frac{a\varphi '}{{\cal H}
\sqrt{1-3{\cal K}\displaystyle\frac{1-c_{_{\rm S}}^2}{n(n+2)}}}\, .
\label{defz}
\end{equation}
This equation was obtained previously in Ref.~\cite{HN}.  The same
remark as for the $u$ equation applies: in the ${\cal K}\neq 0$ the
effective potential is not only $z''/z$ (as in the flat case) but is
corrected by the $+3{\cal K}(1-c_{_{\rm S}}^2)$ term. In addition, the
effective potential now depends on the wave number $n$ through the
quantity $z$. This has to be contrasted with the effective potential
for $u$, $\theta ''/\theta $, which is $n$-independent.

Having defined the various quantities needed to study the evolution of
cosmological perturbations through a bouncing phase, we now turn to
the description of the bounce itself.

\section{Modeling the bounce}

In this section, we define precisely the behavior of the scale factor
during the bouncing epoch, then discuss its relation with the
following eras of standard cosmology and derive the relevant
perturbation potentials.

\subsection{The de Sitter-like bounce}

\begin{figure}[t]
\begin{center}
\includegraphics[width=9cm]{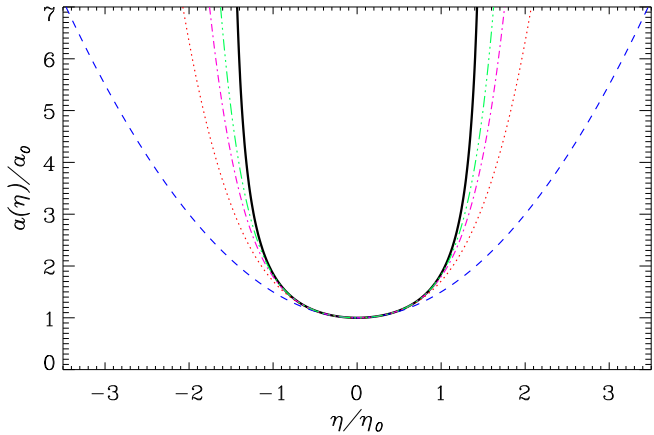}
\caption{Scale factors as functions of the conformal time $\eta$
corresponding to the de Sitter-like solution [Eq.~(\ref{dS}), full
line] and its various levels of approximations stemming from
Eq.~(\ref{aseries}), namely up to quadratic (dashed), quartic
(dotted), sixth (dot-dashed) and eighth power (dot-dot-dashed). The
last two approximations, although clearly better from the point of
view of the scale factor, do not lead to any new qualitative
information as far as the evolution of the perturbations is
concerned.} \label{scales}
\end{center}
\end{figure}

Once the background is fixed, the effective potentials for the
quantities $u$ and $v$ are completely specified. In this section, our
aim is therefore to discuss how one can model the scale factor of a
bouncing universe. At this point, one should notice the differences
(and similarities) with inflation. In an inflationary universe, the
behavior of the scale factor is known: essentially, this is $a\propto
\vert \eta \vert ^{-1}$, \ie the de Sitter phase. However, one can
also treat slightly more complicated backgrounds by means of an
expansion around this de Sitter solution. This expansion is
characterized by the so-called slow-roll parameters~\cite{slowroll},
which are constrained to be small. The de Sitter solution also exists
in the bounce case~\cite{EM} and, as we shall see, it can be used in
much the same way. However, contrary to the inflation case, there is
no fundamental reason why the background equation of state should be
close to vacuum. Despite this fact, one can nevertheless expand around
the $\Ka=1$ de Sitter spacetime and similarly define parameters which
control the departure from it. Obviously, those parameters are not
subject to tight constraints, and in particular are not required to be
small.

\begin{figure}[t]
\begin{center}
\includegraphics[width=9cm]{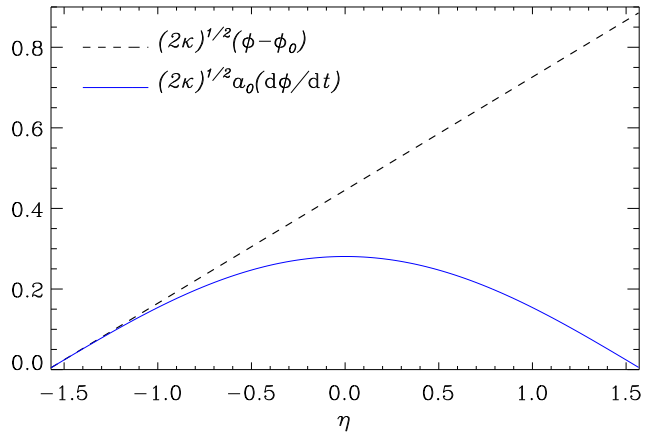}
\caption{Behavior of the scalar field and its coordinate time
derivative as functions of the conformal time $\eta$ (varying between
$-\pi/2$ and $\pi/2$ for the overall evolution of the Universe) for
the de Sitter-like solution with $\eta_0=1.01$.}
\label{phiphop}
\end{center}
\end{figure}

\begin{figure}[t]
\begin{center}
\includegraphics[width=9cm]{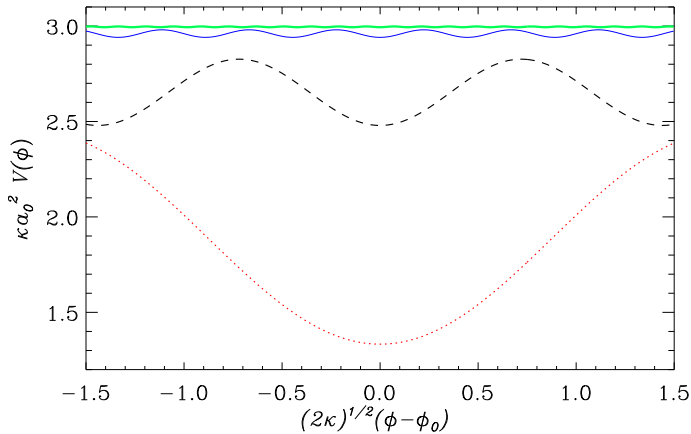}
\caption{The shape (\ref{potphi}) of the potential for the scalar
field $\varphi$ (in units of the Planck mass
$\kappa^{-1/2}=\mP/\sqrt{8\pi}$) for different values of the bounce
characteristic conformal time $\eta_0$. The full lines are
respectively for $\eta_0=1.001$ (above) and $\eta_0=1.01$ (below), the
dashed line corresponds to $\eta_0=1.1$, and the dotted line is for
$\eta_0=1.5$. In the strict de Sitter limit $\eta_0\to 1$, the
potential goes to the constant value $V(\varphi)=3/(\kappa a_0^2)$,
which explains why the $\eta_0=1.001$ seems almost constant as it
oscillates with a very small amplitude around its central value
[($3-1/\eta_0^2$) in these units].}
\label{Vphi}
\end{center}
\end{figure}

For ${\cal K}>0$, the de Sitter solution~\cite{EM} corresponds to the
scale factor $a(t)= a_0 \cosh (\omega t)$, which is expressed as a
function of the cosmic time $t$, with $\omega=1/a_0$. More general
solutions are obtained by relaxing this last constraint and considering
a general value for $\omega$. These de Sitter-like solutions are the
ones we shall be concerned with in what follows: our expansion will be
based on these solutions. In terms of conformal time, one can
integrate the relation $a\dd \eta = \dd t$ to get
\begin{equation}
a(\eta)=a_0\sqrt{1+\tan ^2\biggl(\frac{\eta }{\eta _0}\biggr)},\label{dS}
\end{equation}
where the conformal time is bounded within the range $-\pi
/2<\eta/\eta_0 <\pi /2$ and the conformal time duration $\eta_0$ is
related to the de Sitter coefficient $\omega$ through
$\eta_0=(a_0\omega)^{-1}$ [the solution (\ref{dS}) is shown in
Fig.~\ref{scales}]. 

In order to understand the dynamics of this solution, one needs to
obtain the evolution of the scalar field. It can be integrated
straightforwardly with the scale factor (\ref{dS}): from
Eqs.~(\ref{rho}) and (\ref{p}), one obtains
\begin{equation}
\varphi = \varphi_0 + \sqrt{\frac{2\Ups}{\kappa}}\left( \eta +
\frac{\pi}{2} \eta_0\right),\label{scalareta}
\end{equation}
where we have set $\varphi\to\varphi_0$ as the cosmic time $t\to
-\infty$, \ie as $\eta/\eta_0\to -\pi/2$, and we also have defined a
parameter
\begin{equation}
\Ups \equiv 1 - \frac{1}{\eta_0^2} \label{Upsilon}
\end{equation}
for further convenience. We shall keep this definition later on for
more general bounces than the quasi-de Sitter ones.

It should be noted that the parameter $\Ups$, in the case of de Sitter
like expansion (\ref{dS}) is, according to the definition (\ref{nec}),
$\Ups_\mathrm{dS}= \Hu^2 \Gamma$, which is proportional to
$\rho+p$. As a result, the null energy condition at the bounce can
only be satisfied provided $\Ups>0$, \ie if $\vert \eta
_0 \vert \ge 1$: indeed, one has
\begin{equation}
\lim_{\eta\to 0}(\rho +p)=2\frac{\Ups}{a_0^2},\label{limnec}
\end{equation}
a relation which we shall use in the rest of the paper to define
$\Ups$ in a solution-independent way. As emphasized before, the case
$\eta _0=1$ corresponds to a constant scalar field potential and to an
equation of state $\rho =-p$ and is thus the exact counterpart of the
inflationary de Sitter solution.  The scalar field time derivative is
now simply obtained as
\begin{equation}
\frac{\dd\varphi}{\dd t}= \frac{\dd\varphi}{a \dd \eta} =
\frac{1}{a_0} \left\{ \frac{2\Ups}{\kappa \left[ 1+ \tan^2 \displaystyle
\left( \frac{\eta }{\eta _0} \right) \right]} \right\}^{1/2}.
\end{equation}
Both the field and its time derivative are displayed in
Fig.~\ref{phiphop} for a case having $\eta_0\not=1$ as functions of
the conformal time. 

It is now a simple matter to derive the corresponding potential for
the scalar field which solves Einstein equations (\ref{rho}) and
(\ref{p}). It reads
\begin{equation}
\kappa a_0^2 V(\varphi
)=\frac{\Hu'+2\left(\Hu^2+\Ka\right)}{\left(a/a_0\right)^2},
\end{equation}
\ie, with the solution (\ref{dS}) above,
\begin{equation}
\kappa a_0^2 V(\varphi )=\frac{3}{\eta _0^2}+2 \Ups \sin
^2\left[\frac{\sqrt{2\kappa}}{\eta _0} \Ups^{-1/2}(\varphi -\varphi
  _0)\right]\, 
, \label{potphi}
\end{equation}
and it is displayed in Fig.~\ref{Vphi} with a specific choice of
initial conditions for the field. From Figs.~\ref{phiphop} and
\ref{Vphi}, one sees that the universe starts at either a maximum or a
minimum of the potential, in both cases with a nonvanishing amount of
kinetic energy in the scalar field $\varphi'(-\pi\eta_0/2) =
\sqrt{2\Ups/\kappa}$.

One remarkable property of the above model is that the effective
potential for density perturbations remains very simple even if $\eta
_0\neq 1$. Indeed, assuming the de Sitter like solution (\ref{dS}) and
plugging it into the form (\ref{deftheta}) yields
\begin{equation}
\theta_\mathrm{dS} = \frac{1}{a_0}
\sqrt{\frac{3}{2\left(\eta_0^2-1\right)}}\sin
\left(\frac{\eta}{\eta_0}\right),
\end{equation}
which, together with $\cS^2=-1/3$ [this stems from Eq.~(\ref{cs2})
with the solution (\ref{dS}) and the scalar field (\ref{scalareta})]
leads to
\begin{equation}
V_u^{(\mathrm{dS})} = 4-\frac{1}{\eta_0^2}, \label{dSVu}
\end{equation}
or, in other words, the potential for the variable $u$ does not depend
on time for the de Sitter-like solution. Besides, the maximum value
achievable by this potential is given for the limiting case $\eta_0\to
1$, and it is $V_u^{(\mathrm{dS}, \mathrm{max})}=3$: this potential
can only interact with the modes $n=0$ and $n=1$, which we already
mentioned are gauge modes. This was to be expected since the exact de
Sitter solution, in this bouncing situation as in the more usual
inflationary scenario, does not amplify scalar perturbations by any
amount. We believe that this is what happens in Fig.~4 of
Ref.~\cite{GT} in which the Bardeen potential mode $n=10$ is seen to
oscillate while passing through a de Sitter-like bounce, reflecting
the nondomination of the potential at this point.

\subsection{General bouncing scale factor}

We now assume that the universe experiences a regular bounce at the
time $\eta=0$. This means there exists a particular function $a(\eta)$
which can always be Taylor expanded in the vicinity of $\eta =0$.
Since we are interested in understanding the behavior of the
perturbations through the bounce, and because the effective potentials
for density perturbations involve derivatives of the scale factor only
up to the fourth order, a description of the scale factor up to $\eta
^4$ only is sufficient. We therefore set
\begin{equation}
a(\eta )=a_0\biggl[1+\frac{1}{2}\biggl(\frac{\eta }{\eta _0}\biggr)^2
+\delta \biggl(\frac{\eta }{\eta _0}\biggr)^3+\frac{5}{24}(1+\xi )
\biggl(\frac{\eta }{\eta _0}\biggr)^4 \biggr]\, ,
\label{aseries}
\end{equation}
which defines the parameters $a_0$, the radius of the universe at the
bounce, $\eta_0$, the typical conformal time scale of the bounce,
$\delta$ and $\xi$. They control the magnitude of each term of the
expansion.

For the scale factor (\ref{aseries}) to be a solution of Einstein
equations with a scalar field as matter content, the function
$a(\eta)$ must be chosen such that $\eta_0$ is greater than unity,
which is only a necessary condition as discussed below. The parameters
$a_0$ and $\eta_0$ also provide the tangent de Sitter-like
solution~(\ref{dS}), whereas $\delta$ and $\xi$ measure the deviation
with respect to this de Sitter-like solution.  Eq.~(\ref{aseries})
represents a double expansion, both around $\eta=0$ and around the de
Sitter-like solution discussed in the previous section since $\delta =
\xi = 0$ exactly corresponds to the small $\eta$ expansion of the
scale factor (\ref{dS}); this explains, among others, the factor
$5/24$ in this equation. The parameters $\delta$ and $\xi$ are in a
certain sense similar to the traditional slow-roll parameters. Both
the de Sitter form and its approximations are shown in
Fig.~\ref{scales} as functions of the conformal time.

We now discuss how the general expansion~(\ref{aseries}) can be
related to an underlying particle physics model. The scale factor is
entirely specified once the scalar field potential $V(\varphi )$ and
the initial conditions, for instance the values of the scalar field
and its first derivative at the bounce: $\varphi _0$, $\varphi _0'$,
have been chosen. This means that there exists a relation between
these last quantities and the parameters $a_0$, $\eta _0$, $\delta $
and $\xi $ characterizing the expansion~(\ref{aseries}). We now
establish what this relation is. This can be done easily by solving
the Einstein equations~(\ref{rho}) and (\ref{p}) in the vicinity of
the bounce. In practice, we insert the Taylor expansion of the scalar
field,
\begin{equation}
\varphi (\eta )=\varphi _0+\varphi _0'\eta +\frac12\varphi _0''\eta ^2 
+{\cal O}(\eta ^3)\, ,
\end{equation}
and of the scale factor, Eq.~(\ref{aseries}), in Eqs.~(\ref{rho}) and
(\ref{p}) and identify, order by order, the various terms appearing in
the resulting expressions. To zeroth order, this gives
\begin{equation}
\label{a0}
a_0^2=\frac{6-\kappa \varphi _0'^2}{2\kappa V(\varphi _0)}\, , 
\quad 
\eta _0^2=\biggl(1-\frac{\kappa }{2}\varphi _0'^2\biggr)^{-1}\, .
\end{equation}
The last relation can also be re-written as 
\begin{equation}
\Ups =\frac{\kappa }{2}\varphi _0'^2\, ,
\end{equation}
as expected from Eqs.~(\ref{nec}) and (\ref{limnec}). We see that the
magnitude of the scalar field conformal time gradient at the bounce
determines the value of the parameter $\Ups$. Typically, one expects
$\Ups \ll 1$ since the order of magnitude of the scalar field and its
derivatives should be such that $\varphi _0'\ll m_{\rm Pl}$ in order
for the field theory to make sense. If the velocity of the field
vanishes at the bounce, then $\Ups =0$. 

To first order, the Einstein equations yield
\begin{eqnarray}
\kappa \varphi _0'\biggl(\varphi _0''+a_0^2\frac{{\rm d}V}{{\rm d}\varphi}
\biggl \vert _{\varphi _0}\biggr) &=& 0\, , 
\\
\kappa \varphi _0'\biggl(\varphi _0''-a_0^2\frac{{\rm d}V}{{\rm d}\varphi}
\biggl \vert _{\varphi _0}\biggr) +\frac{12\delta }{\eta _0^3} &=& 0\, .
\end{eqnarray}
In the following, we will be mainly interested in the situation where
the bounce is symmetric, that is, we shall demand that $\delta
=0$. Then, there are two ways of satisfying the Einstein
equations. Either the kinetic energy vanishes at the bounce or
$\varphi _0'\neq 0$ but then $\varphi _0''=0$ and ${\rm d}V/{\rm
d}\varphi \vert _{\varphi _0}=0$. This means that the bounce occurs at
the minimum of the scalar field potential. This also implies that, in
this case, the minimum of the potential cannot vanish, $V(\varphi
_0)\neq 0$, see Eq.~(\ref{a0}). If, for instance, the potential is
given by $V(\varphi ) \propto \varphi ^n$, as is the case for instance
of the model studied in Ref.~\cite{GT}, then the only way to have a
symmetric bounce is to satisfy the condition $\varphi _0'=0$ at the
bounce and, as a consequence, one necessarily has $\Ups =0$. In the
following, we will be mainly interested in the second situation, \ie
$\varphi _0'\neq 0$, since we will show that amplitude of the spectrum
is controlled by the parameter $\Ups $. In this case, going to the
next order allows us to determine what the parameter $\xi $ is. The
result reads
\begin{eqnarray}
\displaystyle \xi &=& 
\frac{1}{5(2-\kappa \varphi _0'^2)^2\kappa V^2(\varphi _0)}\biggl\{
(2-\kappa \varphi _0'^2)V_0\biggl[(6-\kappa \varphi _0'^2)^2
\nonumber \\
& & +6(-2+\kappa \varphi _0'^2)\kappa V(\varphi _0)\biggl]
+(6-\kappa \varphi _0'^2)^2\varphi _0'^2\frac{{\rm d}^2V}{{\rm d}\varphi ^2}
\biggl \vert _{\varphi _0}\biggl\}\, .\nonumber 
\\
& & 
\end{eqnarray}
Let us notice that this parameter depends on the second order
derivative of the potential at the bounce.

Assuming a symmetric bounce for now on, \ie setting $\delta =0$, some
restrictions can be put on the numerical value of $\xi$. They stem
from the fact that we demand $a(\eta)$ to be positive in the range
$-\eta_0 \leq \eta \leq \eta_0$ and to describe a bounce, \ie $a'>0$
for $0 < \eta < \eta_0$. This latter condition turns out to be more
stringent and implies $\xi > -11/5$. Moreover, if we further require
that the scale factor (\ref{aseries}) be solution of Einstein equation
sourced by a single scalar field, we see from Eq.~(\ref{nec}) that
$\Hu^2 \Gamma$ must be positive. Around the bounce, this is
\begin{equation}
\Hu^2\Gamma \simeq \Ups - \frac{5}{2} \xi \left( 1 -2\Ups \right)
\left( \frac{\eta}{\eta_0}\right)^2 +
{\cal O}\left[ \left( \frac{\eta}{\eta_0}\right)^4 \right],
\end{equation}
which will be positive definite in a small but finite neighborhood of
$\eta=0$ provided $\xi<0$ in the limit $\Ups\to 0$ we will be
concerned with. Combining both constraints, we arrive at
\begin{equation}
-\frac{11}{5} < \xi < 0.
\end{equation}
The approximation method we discuss later does not allow to consider
very small values for $\xi$, so that in practice, we shall use
$-11/5<\xi\alt -0.1$.

We now assume the fiducial expansion (\ref{aseries}) for the bounce
through which we want to propagate the perturbations. Let us however
first examine the connection of this bounce to the standard
cosmological epochs of radiation and matter domination.

\subsection{Connecting the bounce to standard cosmology}

In this section, we study how the bounce that we described previously
can be connected to an epoch of the standard hot big bang model. In
particular, we study the connection with a radiation dominated era. In
this case, the scale factor can be written as
\begin{equation}
\label{scalerad}
a(\eta )=a_{\rm r}\sin (\eta -\eta _{\rm r})\, ,
\end{equation}
where $a_{\rm r}$ and $\eta _{\rm r}$ are two parameters to be fixed
with the help of the matching conditions. We match this scale factor
to the bouncing scale factor given by Eq.~(\ref{aseries}), using the
junction conditions, known to be valid even in the curved spatial
section case, as derived in Ref.~\cite{MS}, namely $[a]=[a']=0$. The
matching is performed at $\eta =\eta _{\rm j}$ such that $\eta _{\rm
j}\ll \eta _0$ in order for our quartic approximation of the scale
factor to still be meaningful. The matching conditions imply that the
Hubble parameter at the matching time is given by
\begin{equation}
{\cal H}(\eta _{\rm j})=\frac{x}{\eta
_0}\frac{1+\displaystyle\frac{5}{6}\left(1+\xi
\right)x^2}{1+\displaystyle\frac{1}{2}x^2+\displaystyle\frac{5}{24}
\left( 1+\xi \right) x^4}\, ,
\end{equation}
where $x\equiv \eta _{\rm j}/\eta _0\ll 1$. From the above formula,
one sees that it is not possible to connect the bounce to an epoch
where the curvature is negligible, provided the null energy condition,
which demands $\eta _0\ge 1$ [see discussion around
Eq.~(\ref{limnec})], is still satisfied. As a consequence, this implies
that ${\cal H}(\eta _{\rm j})$ cannot be large in comparison to ${\cal
K}=1$; in fact, since $\eta_0\simeq 1$ and $x\ll 1$, $\Hu^2$ is
expected to be negligibly small compared to unity right after the
bounce. This means that one necessarily connects the bounce to a
regime where the curvature is important or, in other words, in a
region where the sine function appearing in the scale
factor~(\ref{scalerad}) cannot be approximated by the first term of
the Taylor expansion, $a(\eta ) \simeq a_{\rm r}(\eta -\eta _{\rm
r})$. The only way to avoid this conclusion would be to violate the
null energy condition, as already noticed in Ref.~\cite{ppnpn1} and to
have a small $\eta _0$ but then it would have been useless to consider
the case ${\cal K}=1$ for modeling the bounce since this was done
precisely in order to satisfy this condition. Therefore, we conclude
that between the bounce and the standard hot big bang, another phase
must necessary occur whose main effect will be to drive ${\cal H}$ to
sufficiently large values. This is usually the role played by a phase
of inflation.

With the general framework thus clarified, let us turn to the
evolution of the scalar gravitational perturbations through the bounce
by means of evaluating the effective potential for the variable $u$
related with the Bardeen potential through Eq.~(\ref{defu}). We
discuss the potential for the variable $v$ in the discussion section
\ref{uetv} below.

\subsection{The potential $V_u(\eta )$}

\begin{figure*}
\begin{center}
\includegraphics[width=8.5cm]{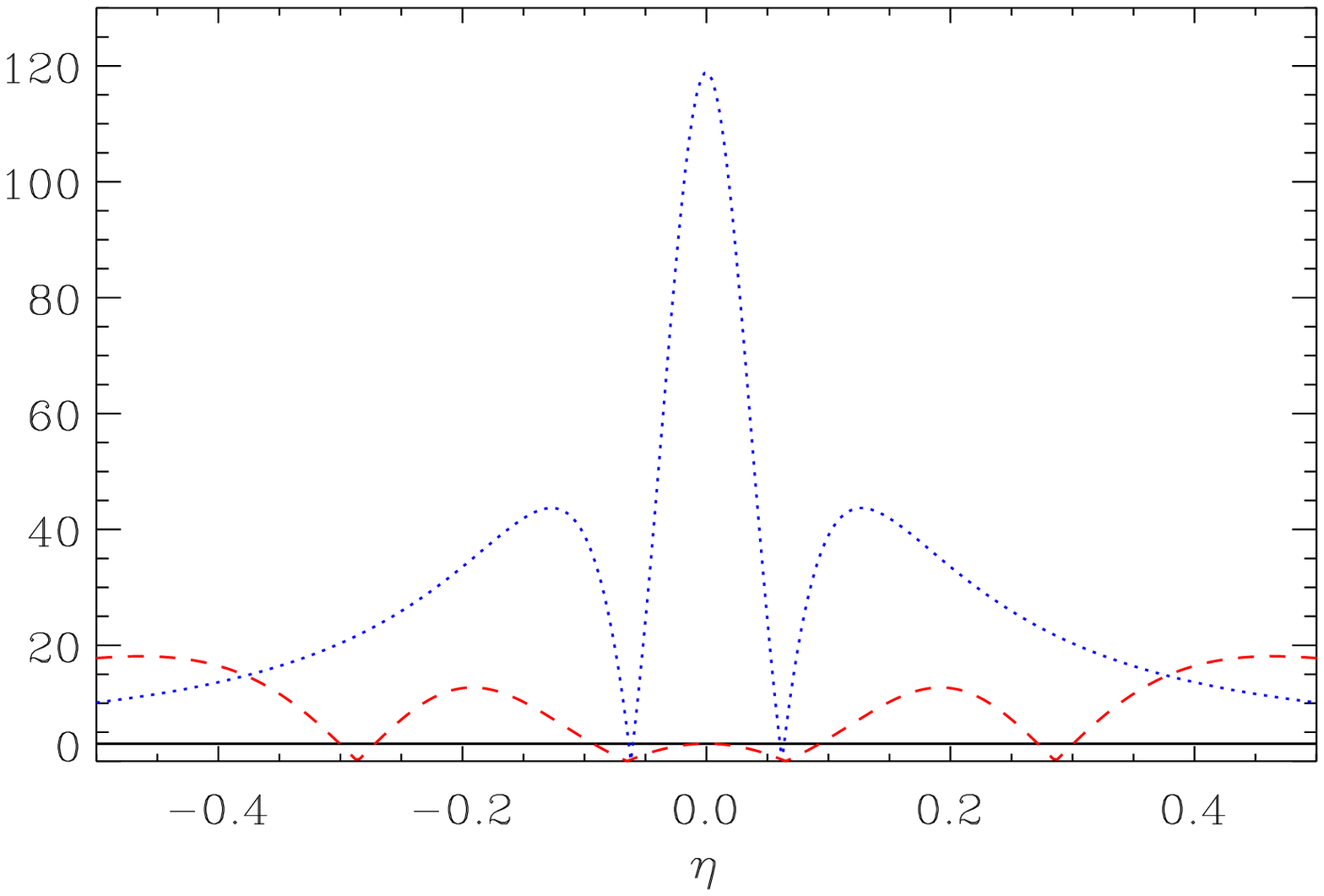}
\includegraphics[width=8.5cm]{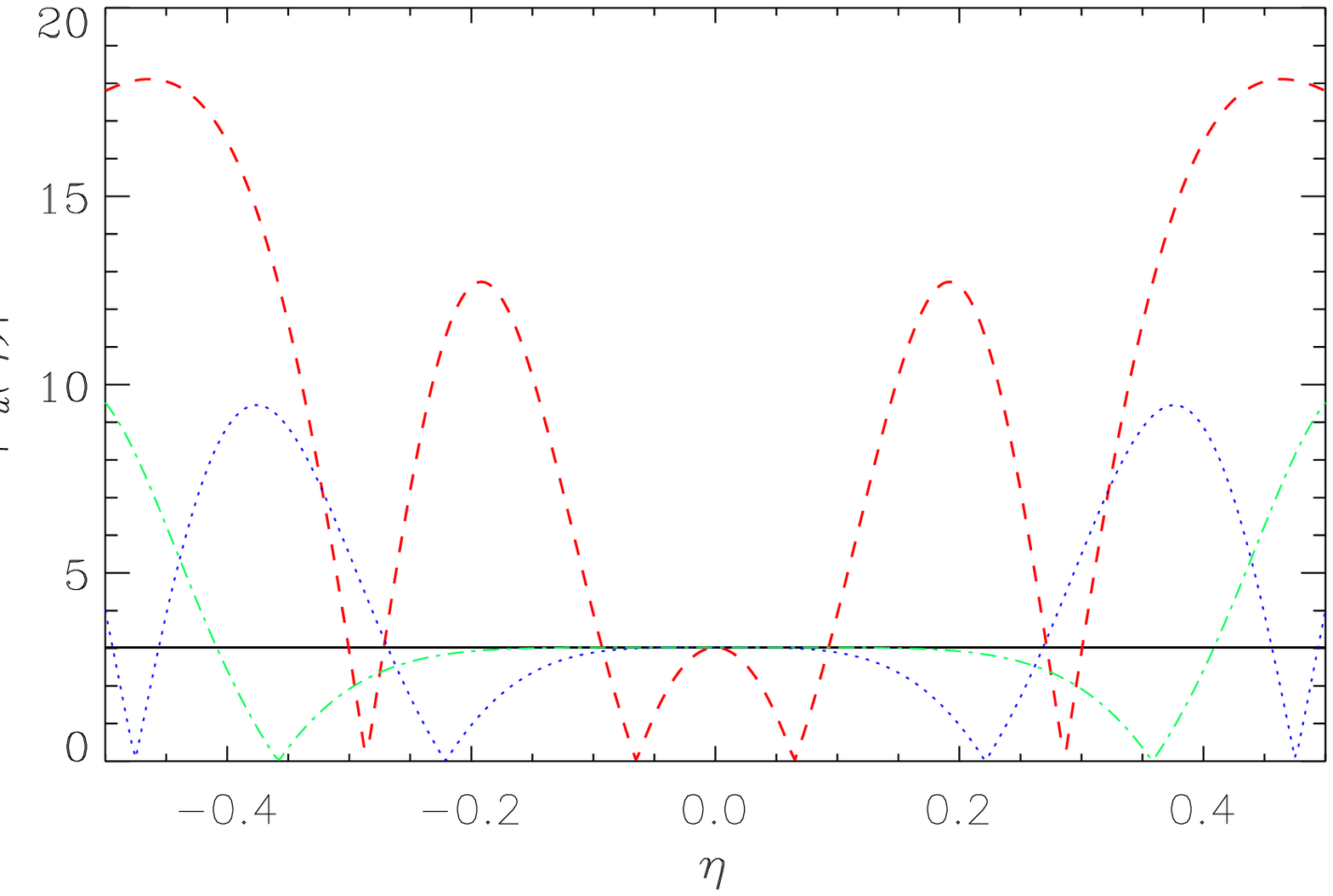}
\caption{Absolute value of the effective potential $V_u(\eta)$ for the
perturbation variable $u(\eta)$ for the de Sitter-like case (full line
on both panels), for which it is constant and for the various
approximation levels (from quadratic to eighth power of the scale
factor). The left panel shows the potential as obtained by using the
quadratic (dotted line) and quartic (dashed) expansions of the scale
factor only, whereas the right panel presents the situation when
quartic (dashed), sixth (dotted) and eighth (dot-dashed) terms are
used. It is clear that the quadratic approximation is qualitatively
wrong and cannot be used to describe a de Sitter bounce. The value
$\eta_0=1.01$ has been used to derive these plots.}
\label{dSpot}
\end{center}
\end{figure*}

\begin{figure}
\begin{center}
\includegraphics[width=9cm]{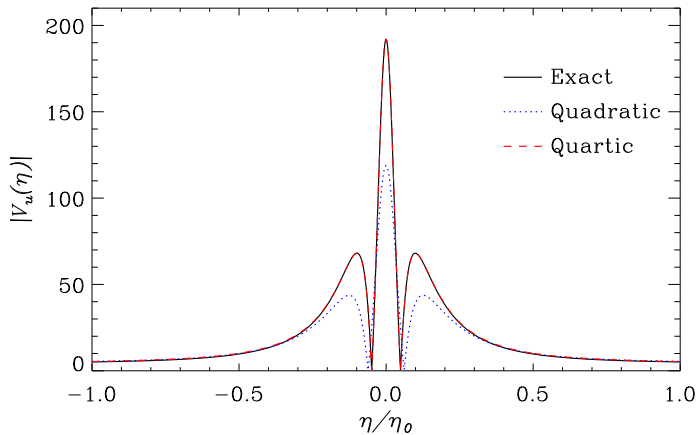}
\caption{Absolute value of the potential $V_u(\eta)$ as a function of
rescaled conformal time $\eta/\eta_0$ for $\eta_0=1.01$ as derived
using either the assumption that the scale factor behaves as a square
root, \ie $a=a_0\sqrt{1+\left( \eta/\eta_0 \right)^2 }$, (full line)
or Eq.~(\ref{aseries}) up to quadratic (dotted line) and quartic order
with $\delta=0$ and $\xi=-2/5$ (dashed line). The quartic
approximation is extremely close to the exact solution, exemplifying
its accuracy, while the quadratic approximation appears to be at best
qualitatively correct.}
\label{Vu24}
\end{center}
\end{figure}

The effective potential for the variable $u$ in the de Sitter-like
solution is, according to Eq.~(\ref{dSVu}), constant in time. This is
however very specific to this particular solution, as any displacement
away from it immediately leads to a different form of the
potential. This is illustrated in Fig.~\ref{dSpot} which shows the
relative accuracy of the expansion (\ref{aseries}) around the de
Sitter-like solution (\ref{dS}). It is also clear from the figure that
the expansion (\ref{aseries}), if pushed to sufficiently high orders
in $\eta$, gives back the correct constant value over a large range of
conformal times. Let us now turn to the more general bounce case of
Eq.~(\ref{aseries}).

\begin{figure}[t]
\begin{center}
\includegraphics[width=9cm]{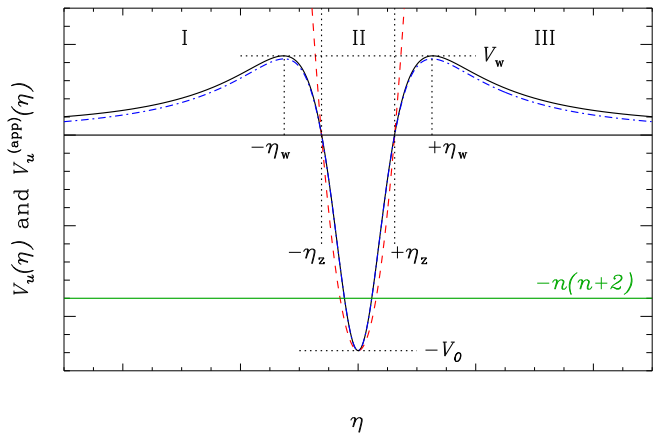}
\caption{The potential $V_u (\eta)$ [full line, Eq.~(\ref{Vureal})]
  and its approximation stemming from Eq.~(\ref{approxVu})
  $V_u^\mathrm{(app)} (\eta)$ (dot-dashed line) in full details with
  the same parameters as in Fig~\ref{Vu24}. Also shown is the
  parabolic approximation that will be used in Sec.~\ref{piecewise}
  (dashed line), the various conformal times involved in the
  calculations in the text, as well as a pictorial definition of the
  regions I, II and III used for the matching of the perturbations
  also in Sec.~\ref{piecewise}. The mode $n(n+2)$ interacts, in this
  example, only with the central part of the potential.}
\label{fitPRD}
\end{center}
\end{figure}

Arbitrary values for the parameter $\xi$ restricted to the range of
interest discussed above lead to the generic shape illustrated in
Fig.~\ref{Vu24}. The calculation of the effective potential is
extremely complicated even with the quartic approximation of the scale
factor. Even if it can be done in full generality since, for a scale
factor given by Eq.~(\ref{aseries}), the potential $V_u(\eta)$ reads
\begin{equation}
V_u(\eta )\equiv \frac{\theta''}{\theta} + 3\Ka \left( 1-\cs^2\right)
=\frac{P_{24}(\eta )}{Q_{24}(\eta )}\, ,\label{Vureal}
\end{equation}
where $P_{24}(\eta )$ and $Q_{24}(\eta )$ are two polynomials of order
$24$, in practice the calculation is not tractable. However, since in
practice we always have $\eta /\eta _0\ll 1$, only the first monomials
are important. One can check that the following approximation
\begin{equation}
\label{approxVu}
V_u^{\rm (app)}(\eta )=3\frac{c_0+c_2\eta ^2}{d_0+d_2\eta ^2+d_4\eta
^4}\, ,
\end{equation}
is extremely good, see Fig.~\ref{fitPRD}. In this expression, we have
only kept the first two monomials at the numerator and the first three
ones at the denominator.
The coefficients $c_i$'s and $d_i$'s can be written as 
\begin{eqnarray}
c_0 &=& 2\eta _0^2(\eta _0^2-1)(2-10\eta _0^2+8\eta _0^4+5\xi )\, ,
\\
c_2 &=& 49 +48\eta _0^6+55\xi +50\xi ^2 -18\eta _0^4(6+5\xi )
\nonumber \\
& & +\eta _0^2(11+35\xi )\, ,
\\
d_0 &=& 12\eta _0^4(\eta _0^2-1)^2\, 
\\
d_2 &=& 12\eta _0^2(\eta _0^2-1)(-3+3\eta _0^2-5\xi )\, 
\\
d_4 &=& -1+15 \eta _0^4(4+\xi )+5\xi (16+15\xi )-\eta _0^2(59+95\xi )\, .
\nonumber 
\\
\end{eqnarray}
Equipped with Eq.~(\ref{approxVu}), we can now compute the height and
the position of the central peak and of the wings. Let us start with
the central peak. The absolute value of $V_u(\eta )$ at $\eta =0$ is
given by
\begin{equation}
V_0= \frac{3c_0}{d_0}=\frac{2-10\eta _0^2+8\eta _0^4
+5\xi }{2\eta _0^2(\eta _0^2-1)}\, . \label{V0tot}
\end{equation}
The most important property of the above formula is that it diverges
as $\eta _0 \rightarrow 1$. It is shown in Ref.~\cite{MP} that this
property also holds in the case $\delta\not= 0$, and is therefore
generic, \ie not restricted to symmetric bounces. We have seen
previously that the values of $n$ of astrophysical interest are such
that $n\gg 1$. Therefore, a necessary condition for the bounce to
affect the spectrum of the fluctuations is that $\eta _0$ be close to
one. As already discussed, the physical interpretation is that one
must be very close to a violation of the null energy condition. In
this case, it is more convenient to work with the variable $\Ups $
introduced in Eq.~(\ref{Upsilon}). In practice, $\Ups$ must be a tiny
number in order to get a modification of the spectrum. We have seen
that this is to be expected since $\Ups$ is the square of the ratio of
the scalar field conformal time gradient at the bounce to the Planck
mass. The amplitude of $\Ups$ controls the maximum value of $n$ below
which the perturbation modes will be affected by the bounce. The
crucial point is that for $\Ups\ll 1$, large scales, having
cosmological and astrophysical relevance, can be modified as they
evolve through the bounce.

Assuming for now on that $\Ups\ll 1$, it is sufficient to Taylor
expand everything in terms of this parameter to get an accurate
approximation. For $V_0$, one gets
\begin{equation}
V_0=-\frac{5\xi }{2\Ups }-(3-5\xi )+{\cal O}(\Ups)\, .
\label{V0Ups}
\end{equation}
Another interesting quantity is the time $\eta _{\rm z}$ for which
$V_u=0$, see Fig.~\ref{fitPRD}. This time is given by $\eta ^2_{\rm
z}=-c_0/c_2$ which leads to
\begin{equation}
\eta _{\rm z}=\sqrt{-\frac{\Ups}{5\xi }}+{\cal O}(\Ups ^{3/2})\, .
\label{etaz}
\end{equation}
The above equations means that, in the limit $\Ups \rightarrow 0$, the
width of the potential goes to zero while its height increases
unboundedly. Finally, let us describe the wings of the potential. The
position of the wings can be derived from the condition $V_u'=0$
($\eta \neq 0$). This gives
\begin{equation}
\eta _{\rm w}^2=-\frac{1}{c_2d_4}\biggl[c_0d_4\pm
\sqrt{c_2d_4(c_2d_0-c_0d_2)+c_0^2d_4^2}\biggr]\, ,
\end{equation} 
and the Taylor expansion in $\Ups $ reads
\begin{equation}
\eta _{\rm w}=\sqrt{-\frac{4\Ups}{5\xi }}+{\cal O}(\Ups ^{3/2})\, .
\label{etaw}
\end{equation}
One sees that $\eta _{\rm w}\simeq 2\eta _{\rm z}$ at first order in
$\Ups $. Therefore, $\eta _{\rm w}$ also goes to zero when $\Ups $
tends to zero. The height of the wings is just given by $V_u(\eta
_{\rm w})$ and can be expressed as
\begin{equation}
V_{\rm w}=-\frac{5\xi }{6\Ups }+\left(3+\frac{5}{3}\xi \right)
+{\cal O}(\Ups )\, .
\end{equation}
The height of the wing also diverges as one approaches the violation
of the null energy condition and, at first order in $\Ups $, one has
$V_0/V_{\rm w}\simeq 3$. This concludes the description of the
perturbation potential with which we now examine the fate of the
perturbations themselves.

\section{Calculation of the transfer matrix}

The purpose of this section, which is also our main result, is to show
that the transfer matrix $\mathbf{T}$ of Eq.~(\ref{transT}) may depend
on the wave number $n$ in a way which we derive. We found that two
completely different and independent methods, one based on a piecewise
expansion of the potential and the other assuming the potential to
behave mathematically as a distribution rather than a simple function
in the limit $\Ups\to 0$, lead to comparable results, both for the
final spectrum itself and its magnitude. We examine these methods in
turn.

\vspace{0.3cm}

\subsection{Method I: piecewise solution approach} \label{piecewise}

When trying to evaluate the transfer matrix of Eq.~(\ref{transT}) with
the potential $V_u (\eta)$ derived in the previous section, one
immediately faces a difficulty, namely that unfortunately, even with
the simple form given by Eq.~(\ref{approxVu}), the equation of motion
of the variable $u$ is not integrable analytically. However, one can
find piecewise solutions. For a mode which interacts only with the
central peak of the barrier, \ie  which is above the wings, the
potential is essentially zero for $\vert \eta \vert >\eta _{\rm
z}$. This corresponds to regions I and III in Fig.~\ref{fitPRD}. In the
central region, region II in Fig.~\ref{fitPRD} corresponding to $\vert
\eta \vert <\eta _{\rm z}$, we model the bounce by a parabola with a
minimum at $-V_0$ and which vanishes at $\eta =\pm \eta _{\rm z}$. To
summarize, our piecewise potential is given by
\begin{equation}
V_u(\eta )= 
\begin{cases}
0\, ,& \eta < -\eta _{\rm z} \, ,\cr 
\displaystyle -V_0\biggl[1
-\biggl(\frac{\eta }{\eta _{\rm z}}\biggr)^2
\biggr]\, , & -\eta _{\rm z}<\eta <\eta _{\rm z}\,   , \cr 
0\, , & \eta >\eta _{\rm z}.
\end{cases} 
\end{equation}
In each region, the function $u$ is the sum of two modes and 
can be expressed as 
\begin{equation}
u_i(n,\eta )=A_i(n)f_i(n,\eta )+B_i(n)g_i(n,\eta )\, ,
\quad i={\rm I,II,III}\,.
\end{equation}
Before and after the interaction with the barrier, the solution are
plane waves,
\begin{equation}
f_{_{\rm I,III}}(\eta )=\frac{1}{\sqrt{2k}}{\rm e}^{-ik\eta }\, ,
\quad g_{_{\rm I,III}}(\eta )=\frac{1}{\sqrt{2k}}{\rm e}^{ik\eta }\, ,
\end{equation}
where we have introduced the quantity $k\equiv \sqrt{n(n+2)}$ and the
normalization is chosen such as to simplify further calculations (unit
Wronskian). In region II, one has an even and an odd mode, \ie
\begin{equation}
f_{_{\rm II}}(-\eta )=f_{_{\rm II}}(\eta )\, ,\quad 
g_{_{\rm II}}(-\eta )=-g_{_{\rm II}}(\eta )\, .
\end{equation}
For the moment, we do not specify what $f_{_{\rm II}}(\eta )$ and
$g_{_{\rm II}}(\eta )$ are since we are trying to keep the calculation
as general as possible (for example, we could imagine other
parametrization of the potential in the central region for which
$f_{_{\rm II}}$ and $g_{_{\rm II}}$ would be different). Our goal is
to predict what $A_{_{\rm III}}$ and $B_{_{\rm III}}$ are. For this
purpose, we match $u$ and its derivative $u'$ at $\eta =\pm \eta _{\rm
z}$. Straightforward calculations lead to
\begin{widetext}
\begin{equation}
\left[
\begin{matrix}
A_{_{\rm III}} \cr \cr B_{_{\rm III}}
\end{matrix}
\right]
=\frac{1}{2ikW(n)}{\rm e}^{2ik\eta _{\rm z}}
\left[
\begin{matrix}
-f'_{_{\rm II}}+ikf_{_{\rm II}} & -g'_{_{\rm II}}+ikg_{_{\rm II}}
\cr \cr
{\rm e}^{-2ik\eta _{\rm z}}(f'_{_{\rm II}}+ikf_{_{\rm II}}) 
&
{\rm e}^{-2ik\eta _{\rm z}}(g'_{_{\rm II}}+ikg_{_{\rm II}}) 
\end{matrix}
\right] \cdot
\left[
\begin{matrix}
g'_{_{\rm II}}-ikg_{_{\rm II}} & 
{\rm e}^{-2ik\eta _{\rm z}}(g'_{_{\rm II}}+ikg_{_{\rm II}}) 
\cr \cr
f'_{_{\rm II}}-ikf_{_{\rm II}} 
&
{\rm e}^{-2ik\eta _{\rm z}}(f'_{_{\rm II}}+ikf_{_{\rm II}}) 
\end{matrix}
\right]
\left[
\begin{matrix}
A_{_{\rm I}} \cr \cr B_{_{\rm I}}
\end{matrix}
\right]
\, ,
\end{equation}
\end{widetext}
where $W(n)$ is the Wronskian of the function $f_{_{\rm II}}$ and
$g_{_{\rm II}}$, namely $W(n)=f_{_{\rm II}}g'_{_{\rm II}}-f'_{_{\rm
II}}g_{_{\rm II}}$. In the previous expressions, all the functions are
expressed at the point $\eta =\eta _{\rm z}$ (we have used the parity
properties of the function $f_{_{\rm II}}$ and $g_{_{\rm II}}$). The
above matrix is general and is parametrized by only four numbers:
$f_{_{\rm II}}$, $g_{_{\rm II}}$, $f'_{_{\rm II}}$ and $g'_{_{\rm
II}}$. Any model permitting to calculate what these numbers are allows
us to estimate the transfer matrix on the bounce given above. 

We now use the parabolic model introduced before. If we perform the
following change of variable, $\eta \equiv \sqrt{\eta _{\rm
z}/(2\sqrt{V_0})}x$, then the equation of motion for $u$ in region II
takes the form
\begin{equation}
\label{cylineq}
\frac{{\rm d}^2u}{{\rm d}x^2}-\biggl(\frac{x^2}{4}+\alpha \biggr)u=0\, ,
\end{equation}
where the parameter $\alpha $ is given by
\begin{equation}
\alpha =-\frac{1}{2}\eta _{\rm z}\sqrt{V_0}
\biggl[1+\frac{n(n+2)}{V_0}\biggr]\, .
\end{equation}
Equation~(\ref{cylineq}) can be solved exactly in terms of cylinder
parabolic functions~\cite{Grad}. Since the potential is symmetric, the
solutions can always be chosen to be even and odd. The explicit
expression of the even and odd solutions are respectively
\begin{eqnarray}
f_{_{\rm II}}(\eta ) &=& {\rm e}^{-\sqrt{V_0}\eta ^2/(2\eta _{\rm z})}
{}_1F_1\biggl(\frac{\alpha }{2}+\frac14
;\frac{1}{2};\frac{\sqrt{V_0}}{\eta _{\rm z}}\eta ^2\biggr)\, ,
\\
g_{_{\rm II}}(\eta ) &=& \eta 
{\rm e}^{-\sqrt{V_0}\eta ^2/(2\eta _{\rm z})}
{}_1F_1\biggl(\frac{\alpha }{2}+\frac34
;\frac{3}{2};\frac{\sqrt{V_0}}{\eta _{\rm z}}\eta ^2\biggr)\, ,
\end{eqnarray}
where ${}_1F_1$ is the Kummer confluent hypergeometric function. As
already mentioned previously, these functions and their derivatives
must be evaluated at $\eta =\eta _{\rm z}$ and then expanded in the
parameter $\Ups $. The first step is to calculate the parameter
$\alpha $.  This gives
\begin{equation}
\alpha =-\frac{1}{2\sqrt{2}}+\frac{\{-61+5[-53+8n(n+2)]\xi \}}
{200\sqrt{2}\xi ^2}\Ups +{\cal O}(\Ups ^2)\, .
\end{equation}
Using this expansion and that of $V_0$ and $\eta _{\rm z}$, one
obtains at first order in $\Ups $
\begin{widetext}
\begin{eqnarray}
f_{_{\rm II}}(\eta _{\rm z}) &=& {\rm e}^{-1/(2\sqrt{2})}
{}_1F_1\biggl(\frac{2-\sqrt{2}}{8},\frac12,\frac{1}{\sqrt{2}}\biggr)
+{\cal O}(\Ups ^{1/2})\simeq 
0.798+{\cal O}(\Ups ^{1/2})\, ,
\\
g_{_{\rm II}}(\eta _{\rm z}) &=&\frac{1}{\sqrt{-5\xi}}{\rm e}^{-1/(2\sqrt{2})}
{}_1F_1\biggl(\frac{6-\sqrt{2}}{8},\frac32,\frac{1}{\sqrt{2}}\biggr)\Ups^{1/2}
+{\cal O}(\Ups ^{3/2})
\simeq 
\frac{0.422}{\sqrt{-\xi }}\Ups ^{1/2}+{\cal O}(\Ups ^{3/2})\, ,
\\
f'_{_{\rm II}}(\eta _{\rm z}) &=& \frac{\sqrt{-5\xi}}{2}
{\rm e}^{-1/(2\sqrt{2})}
\biggl[-\sqrt{2}
{}_1F_1\biggl(\frac{2-\sqrt{2}}{8},\frac12,\frac{1}{\sqrt{2}}\biggr)
+(\sqrt{2}-1)
{}_1F_1\biggl(\frac{10-\sqrt{2}}{8},\frac32,\frac{1}{\sqrt{2}}\biggr)
\biggr]\frac{1}{\Ups ^{1/2}}+{\cal O}(\Ups ^{1/2})\, ,
\\
&\simeq & -\frac{0.711\sqrt{-\xi}}{\Ups ^{1/2}}+{\cal O}(\Ups ^{1/2})\, ,
\\
g'_{_{\rm II}}(\eta _{\rm z}) &=& \frac16
{\rm e}^{-1/(2\sqrt{2})}
\biggl[(6-3\sqrt{2})
{}_1F_1\biggl(\frac{6-\sqrt{2}}{8},\frac32,\frac{1}{\sqrt{2}}\biggr)
+(3\sqrt{2}-1)
{}_1F_1\biggl(\frac{14-\sqrt{2}}{8},\frac52,\frac{1}{\sqrt{2}}\biggr)
\biggr]+{\cal O}(\Ups ^{1/2})\, ,
\\
&\simeq & 0.878+{\cal O}(\Ups ^{1/2})\, .
\end{eqnarray}
\end{widetext}
The expression for the derivatives can be easily recovered if one uses
the following expression giving the derivative of a Kummer
hypergeometric function, ${}_1F_1'(\alpha, \beta ,z) =(\alpha /\beta
){}_1F_1(\alpha+1, \beta+1,z)$, where a prime in this context means a
derivative with respect to the argument $z$ of the hypergeometric
function.

The next step consists in inserting these relations into the general
form of the transfer matrix and then in expanding the resulting
expression in the parameter $\Ups$. The result reads
\begin{equation}
\label{Tumatch}
T_u\simeq -0.624i \sqrt{\frac{-\xi}{n(n+2)}}
\begin{pmatrix}
1 & 1 \cr
-1 & -1
\end{pmatrix}
\frac{1}{\Ups ^{1/2}}\, .
\end{equation}
Several remarks are in order at this point. First, the formula above
applies only for the modes actually interacting with the potential,
namely those having $n(n+2)\leq V_0$, otherwise, $T_u$ is obviously
the identity. Note also that, in the former case, the transfer matrix
is $n$-dependent. This means that the bounce affects the spectrum and
therefore disproves \apriori any general argument stating that the
spectrum should propagate through the bounce without being
modified~\cite{PBB,DV}. Besides, we see that the amplitude is
divergent as $\Ups$ goes to zero. However, one should remember that we
are not interested in the spectrum of $u$ itself but rather in the
spectrum of the Bardeen potential $\Phi$. The relation between $\Phi$
and $u$, Eq.~(\ref{defu}), together with Eq.~(\ref{limnec}), leads to
the remarkable result that the terms in $\Ups$ cancel out exactly and
that the resulting spectrum is $\Ups$-independent, and thus perfectly
finite even in the $\Ups\to 0$ limit. Finally, the $\xi$ dependence of
the overall amplitude is also predicted by this calculation. As
expected, there is no net effect in the limit $\xi\to 0$ at which the
bounce is effectively de Sitter and thus can amplify no amount of
perturbation. In this last case, the calculation leading to
Eq.~(\ref{Tumatch}) is not accurate enough and should be done at a
higher order in $\Ups$ since the leading order vanishes; one should
then find that the transition matrix is essentially the identity (de
Sitter) plus some correction vanishing in the limit $\Ups\to 0$.

\subsection{Method II: Distributional approach}

We show in this section that the previous result can be understood in
very simple terms and that the result of the previous section can be
reproduced by a back-of-the-envelope calculation. The crucial
observation is that the height of the potential $V_u$ diverges as
$\Ups $ goes to zero while its width shrinks to zero. This suggests
that there is something like a Dirac $\delta $-function at
play.\footnote{Note that even though we send $\Ups\to 0$ in this
section, this is merely a computational artifact allowing an easy
calculation of the effect. The true value of $\Ups$ must be
nonvanishing, although tiny, so the calculation of this section is
accurate only for those modes interacting with the
potential. Therefore, the presence of the Dirac distribution in no way
implies the existence of a singular behavior either of the potential
$V_u(\eta)$ or of the modes $u(\eta)$ themselves as long as $\Ups\neq
0$.} To study this point we calculate the integral of the potential.
One gets
\begin{eqnarray}
\int _{-\infty }^{+\infty }[V_u(\tau )-4]{\rm d}\tau & \simeq &
\int _{-\eta _0}^{+\eta _0}V_u^{\rm (app)}(\tau ){\rm d}\tau
\\ 
&=& \biggl(\frac{-5\pi ^2\xi}{8\Ups }\biggr)^{1/2}+{\cal O}(\Ups ^0)\, . 
\end{eqnarray}
Thus, the potential can be re-written as 
\begin{equation}
V_u(\eta )=-C_{\Ups}\Delta _{\Ups}(\eta )\, ,
\end{equation}
where the constant $C_{\Ups}$ is given by $C_{\Ups}\equiv [-5\pi
^2\xi/(8\Ups) ]^{1/2}$ and where the function $\Delta _{\Ups }(\eta )$ is
a representation of the Dirac $\delta $-function, \ie
\begin{equation}
\lim _{\Ups \rightarrow 0}\Delta _{\Ups }(\eta )=\delta (\eta )\, .
\end{equation}
In a certain sense, the potential $V_u(\eta )$ possesses divergences 
``worst'' than a Dirac $\delta $-function. The equation of motion of 
the quantity $u$ can now be written as
\begin{equation}
u''+[n(n+2)+C_{\Ups }\delta (\eta )]u=0\, ,
\end{equation}
\ie a well-known equation in the context of quantum mechanics. The
matching conditions are $[u]=0$ and $[u']=-C_{\Ups }u(0)$, the last
one coming from an integration of the equation of motion across a thin
shell around $\eta =0$. This reduces to
\begin{eqnarray}
A_{_{\rm III}}+B_{_{\rm III}} &=& A_{_{\rm I}}+B_{_{\rm I}}\, ,
\\
A_{_{\rm III}}-B_{_{\rm III}} &=& A_{_{\rm I}}-B_{_{\rm I}}
-\frac{C_{\Ups }}{i\sqrt{n(n+2)}}(A_{_{\rm I}}+B_{_{\rm I}})\, .
\end{eqnarray}
Straightforward algebraic manipulations lead to the following 
transfer matrix, under the assumption that the second term of the last
equation dominates over the first since $C_{\Ups }\to\infty$ as
$\Ups\to 0$,
\begin{equation}
\label{Tudelta}
T_u=-i \sqrt{\frac{-5\pi^2\xi}{32n(n+2)}}
\begin{pmatrix}
1 & 1 \cr
-1 & -1
\end{pmatrix}
\frac{1}{\Ups ^{1/2}}\, .
\end{equation}
It is interesting to compare Eq.~(\ref{Tumatch}) with
Eq.~(\ref{Tudelta}). The numerical coefficient in the above equation
is $\pi \sqrt{5}/(4\sqrt{2})\simeq 1.242$, to be compared with the
coefficient $0.624$ found in Eq.~(\ref{Tumatch}). The difference is
approximatively a factor $1/2$ in the amplitude. This difference can
be interpreted in the following way. When the matrix transfer is
computed using the matching procedure, one uses the parabola formula
for the potential and one neglects the wings of the potential. The
area of the central part of the potential is given by
\begin{equation}
\int _{-\eta _{\rm z}}^{+\eta _{\rm z}}V_u(\tau ){\rm d}\tau 
=\sqrt{-\frac{100\xi }{45\Ups }} =\frac{4\sqrt{2}}{3\pi
}\int _{-\eta _0}^{+\eta _0}V_u^{\rm (app)}(\tau ){\rm d}\tau \, .
\end{equation}
We see that there is factor $4\sqrt{2}/(3\pi )$ between the area below
the central part and the area below the whole potential including the
wings. Since the matching procedure is sensitive to the central part
only whereas the calculation of the Dirac $\delta $-function is
sensitive to the whole potential, we therefore expect a factor
$4\sqrt{2}/(3\pi )$ between the corresponding two amplitudes. We have
$4\sqrt{2}/(3\pi )\simeq 0.600$ and hence we recover approximatively
the factor $1/2$ mentioned above. The correct amplitude is the one
given by the Dirac $\delta $-function calculation and is $\simeq
1.25$.

\section{Discussion}

We now complete the description of the propagation of perturbations
through a general relativistic bounce by some considerations regarding
the variables $u$ and $v$, the spectrum of tensor modes and a
comparison with other known transitions in cosmology.

\subsection{$u$ versus $v$} \label{uetv}

\begin{figure}
\begin{center}
\includegraphics[width=9cm]{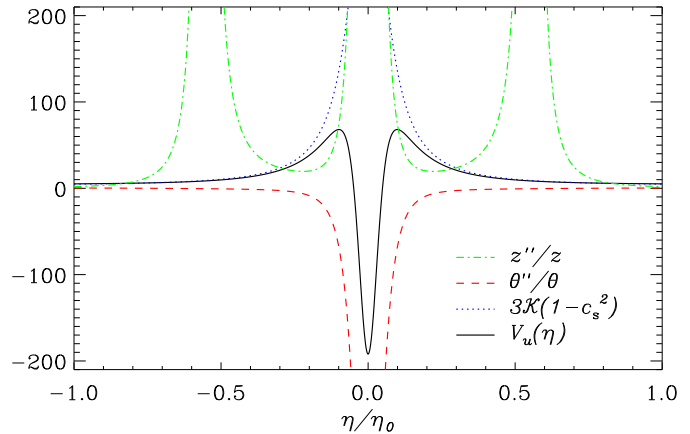}
\caption{Construction of he potentials $V_u$ and $V_v$ for the
perturbation variables $u$ and $v$ in the special case of the square
root form for the scale factor as in Fig.~\ref{Vu24} [or
Eq.~(\ref{aseries}) with $\delta=0$ and $\xi=-2/5$, the corresponding
curves being visually undistinguishable] with $\eta_0=1.01$. According
to Eqs.~(\ref{eomu}) and (\ref{eov}), the potentials depend on three
possible terms, namely $\theta''/\theta$, $3\Ka \left(1-\cs^2\right)$
and $z''/z$, respectively plotted as the dashed, dotted, and
dot-dashed curves. The potential for $u$ is also shown as the full
line. The pole at $\eta=0$ in either $\theta''/\theta$ and $3\Ka
\left(1-\cs^2\right)$ appears with opposite sign but is otherwise the
same, so there is an exact compensation, so that the full potential is
everywhere well-behaved. This is clearly not possible for the
potential $V_v$, since there are more poles in $z''/z$ than there are
in $3\Ka \left(1-\cs^2\right)$, so no compensation can occur at these
points, but also the pole at $\eta=0$ appear with the same sign; the
potential for the function $v$ follows $z''/z$, up to small
corrections and was therefore not plotted here.}
\label{VuVv}
\end{center}
\end{figure}

An interesting issue, debated at length in the
literature~\cite{bounce,ppnpn1,GT,GGV}, is how the variables $u$ and
$v$ behave as they go through the bounce. As a first step towards
understanding what is the variable that is the most useful, let us
construct the potentials for both, as in Fig.~\ref{VuVv}. It is clear
from this figure that the terms appearing in the potential for the
variable $v$, namely
\begin{equation}
V_v (\eta) \equiv \frac{z''}{z} + 3\Ka \left( 1-\cs^2 \right),
\label{Vv}
\end{equation}
never compensate each others, as was found to be the case for $V_u$ of
Eq.~(\ref{Vureal}), so the resulting potential is divergent at some
points which, furthermore, depend on the wavelength index $n$. This
provides a first hint that $v$ is not the correct variable to work
with, and indicates that as one approaches the bounce, or as the
curvature becomes non-negligible, $v$ ceases to be the good quantum
variable (see \eg Ref.~\cite{mfb}).

In the present context, it is easy to show that $u$ and $v$ are
related by the following relation:
\begin{equation}
v=-\frac{1}{\sqrt{1-3{\cal K}\displaystyle\frac{1-c_{_{\rm
S}}^2}{n(n+2)}}} \biggl[u'+\frac{(a\sqrt{\Gamma })'}{a\sqrt{\Gamma
}}u\biggr]\, .
\end{equation}
We know from the previous considerations that $u$ is continuous and
that $u'$ may have a finite jump at $\eta =0$ provided $\Ups$ is small
enough. From the above equation, we conclude that the variable $v$
possesses divergences during the bounce. These divergences are given
by the zeros of the argument of the square root at the denominator of
the previous equation. In other words, $v$ diverges when
\begin{equation}
\label{blowv}
c_{_{\rm S}}^2=1-\frac{n(n+2)}{3},
\end{equation}
as found in Fig.~\ref{VuVv}.

Some remarks are in order at that point. First, an interesting
feature is that $v=0$ at the bounce (hence is regular) and that the
divergences occur before and after the bounce but not at the bounce
itself even though its potential actually diverges at this point. This
is because the effective velocity of sound diverges at the
bounce. Secondly, the time at which $v$ divergences is $n$-dependent
as can clearly be seen from Eq.~(\ref{blowv}). Thirdly, the physical
interpretation of this divergence is subtle. If $v$ had the usual
interpretation (\ie the variable that is canonically quantized), the
divergence would clearly be a problem. Roughly speaking, this would
mean explosive particle creations and, as a consequence that there is
a back-reaction problem. More seriously, this divergence would be at
odd with the fact that the Bardeen potential remains finite and
small. As already discussed below Eq.~(\ref{defv}), there are reasons
to believe that, in the case $\Ka =1$, the variable $v$ introduced
before is not the variable that appears in the action for cosmological
perturbations. This last variable should remain finite during the
bounce.

\subsection{Density perturbations versus gravitational waves}

The evolution of the tensorial modes of perturbations $\mu \equiv a
h$, where $h$ is, roughly speaking, the amplitude of the gravitational
wave, stems from the relation~\cite{MS,grishchuk}
\begin{equation}
\mu''+\left[ n\left(n+2\right) -\Ka -\frac{a''}{a}\right] \mu=0,
\end{equation}
\ie an equation similar to that valid for the scalar modes but with a
potential simply given by $V_h=\Ka + a''/a$. Within the framework of
our bouncing solution, this is
\begin{equation}
V_h = 1+ \displaystyle\frac{1}{\eta_0^2}\frac{1+
  \displaystyle\frac{5}{2}\left(1+\xi\right) \left(
  \displaystyle\frac{\eta}{\eta_0}\right)^2} {1+
  \displaystyle\frac{1}{2}\left(
  \displaystyle\frac{\eta}{\eta_0}\right)^2 +
  \displaystyle\frac{5}{24}\left(1+\xi\right) \left(
  \displaystyle\frac{\eta}{\eta_0}\right)^4},
\label{Vh}
\end{equation}
which can be simply analyzed as follows. 

As direct calculation reveals, the potential $V_h$ of Eq.~(\ref{Vh})
has either a single maximum located at $\eta=0$ if the expansion
parameter $\xi \leq -4/5$, or a minimum at $\eta=0$ and two maxima at
the points $\eta_\mathrm{max}$ given by
\begin{equation}
\eta_\mathrm{max}^2 = \frac{2}{5}\eta_0^2 \frac{\sqrt{5\left(
    5+6\xi\right)} -1 }{1+\xi},
\end{equation}
provided $-6/5\leq \xi \leq -4/5$. In the latter case, the maximum
value attained by the gravitational wave potential is
\begin{equation}
V_h^\mathrm{max} = 1+\frac{1}{\eta_0^2}
\begin{cases}
\displaystyle\frac{15 \left(1+\xi
\right)}{2+\sqrt{5\left(5+6\xi\right)}},& \hbox{ if }
-\displaystyle\frac{6}{5} \leq \xi \leq -\displaystyle\frac{4}{5},\cr
& \cr
1, &\hbox{ otherwise.}
\end{cases}
\end{equation}
Since $\xi <0$ and $\eta_0 >1$, this means that the maximum value for
the potential is less than $22/7\simeq 3.14$ in all cases of physical
interest. In other words, and since in these units the cosmologically
relevant modes are those having $n\gg 1$, the potential is dominated
at all times during the bounce itself, and therefore cannot lead to
tensor mode production. There is therefore a qualitative difference
between the tensor and the scalar modes since the latter can be
affected by the bounce provided the NEC is almost violated, while the
former are never affected, regardless of the underlying parameter
values.

\subsection{Comparison with other transitions}\label{other}

\begin{figure}
\begin{center}
\includegraphics[width=8.5cm]{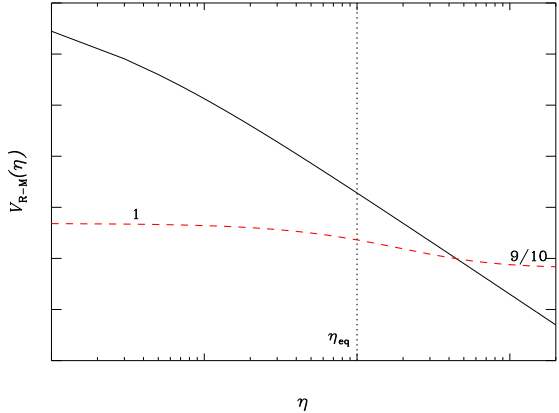}
\caption{The effective potential $V_\mathrm{R-M}(\eta)$ for the
perturbation variables $u(\eta)$ or $v(\eta)$ for the radiation to
matter transition, derived from the scale factor given by
Eq.~(\ref{aradmat}). This log-log sketch shows the potential (full
line) for either of the variables (they differ by numerical factors)
as well as the exact solution (for $k=0$) for the Bardeen potential
(dashed line) as a function of conformal time.}
\label{910}
\end{center}
\end{figure}

In order to make a comparison of our bouncing era to other known
transitions, we first consider below the radiation to matter
transition under the hypothesis that this occurs, as observation
demands, at some time $\eta_\mathrm{eq}$ such that the three-space
curvature is negligible. In other words, we study this transition with
$\Ka=0$. The scale factor can be given the form
\begin{equation}
a(\eta )=a_{\rm eq}\left[b_2^2
\left(\frac{\eta }{\eta _{\rm eq}}\right)^2
+2b_1\left(\frac{\eta }{\eta _{\rm eq}}\right)\right],
\label{aradmat}\end{equation}
where $b_1=b_2=\sqrt{2}-1$ is chosen such that $a(\eta =\eta
_\mathrm{eq}) =a_\mathrm{eq}$. We have emphasized the two different
normalization factors $b_1$ and $b_2$ because they play a different
role in the potentials for either $u$ or $v$. Indeed, the potential
for $v$ in this case is
$\left(a\sqrt{\Gamma}\right)''/\left(a\sqrt{\Gamma}\right)$, which,
for a purely radiation dominated universe (\ie with $b_2=0$ and
$b_1\not=0$), is identically vanishing, whereas the potential for $u$
is $\left[\left(a\sqrt{\Gamma}\right)^{-1}\right]''/\left[
\left(a\sqrt{\Gamma}\right)^{-1}\right]$, which in the same situation
would be $\sim 2/\eta^2$. During the transition however, the presence
of an amount of matter, however tiny, leads to a nonvanishing $b_2$,
and hence a nonzero diverging term for small conformal times $\sim
b_2^2/(2b_1\eta_\mathrm{eq} \eta)$: the radiation dominated universe
represents a singular limiting case. This means that both potentials
are large already at small times, deep into the radiation era, and the
approximation $k^2\ll V_\mathrm{R-M}$ is accurate both before and after
the transition and for both variables. This accounts for the fact that
the Bardeen potential changes during the transition, but only insofar
as the amplitude is concerned, leaving its spectrum unaltered. This is
because, in this situation, there is no potential crossing: the modes
are always below the barrier. Fig.~\ref{910} illustrates this fact and
summarizes the situation by showing a sketch of the perturbation
potential $V_\mathrm{R-M}$ together with the evolution of the
gravitational potential.

\begin{figure*}[t]
\begin{center}
\includegraphics[width=8.5cm]{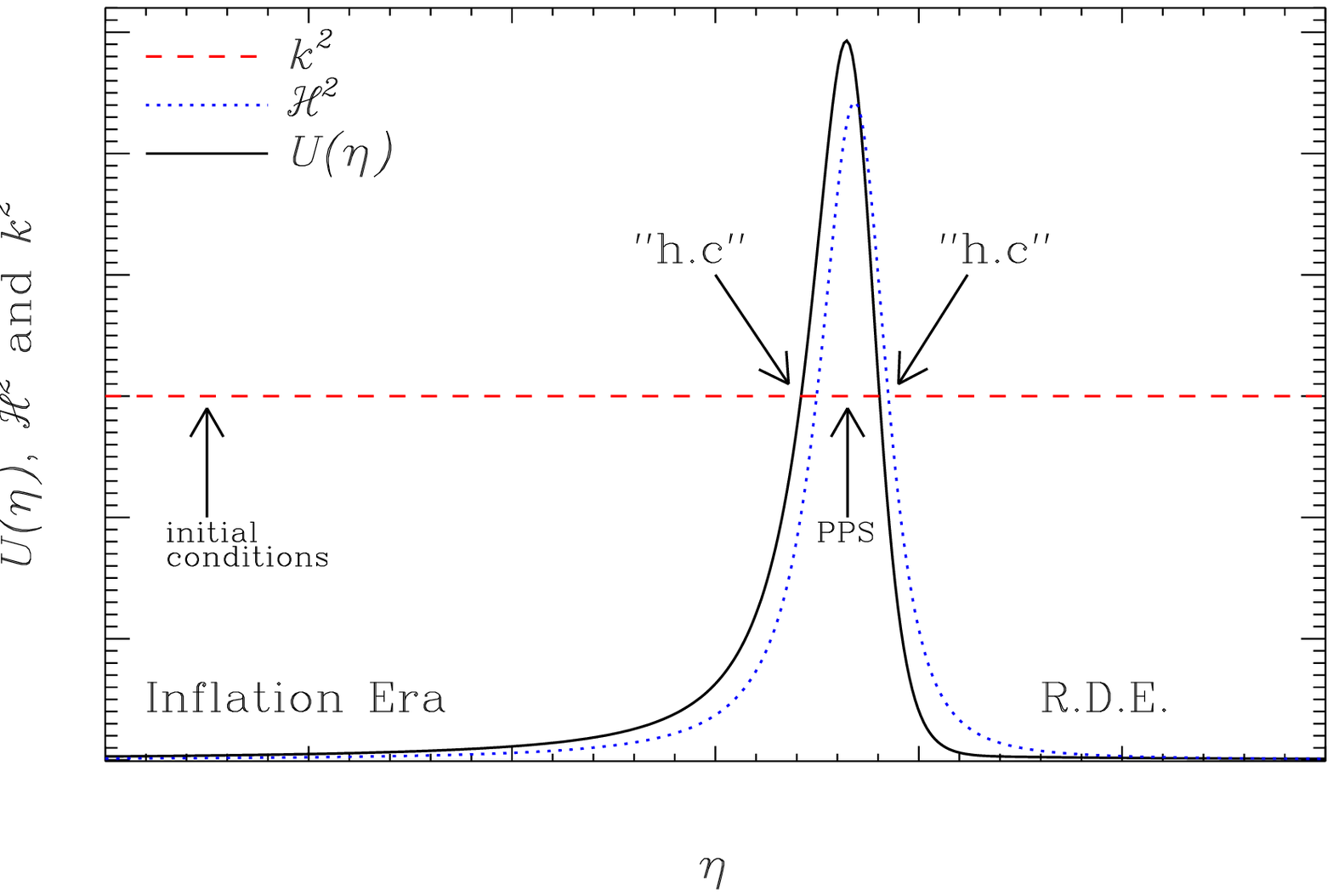}
\includegraphics[width=8.5cm]{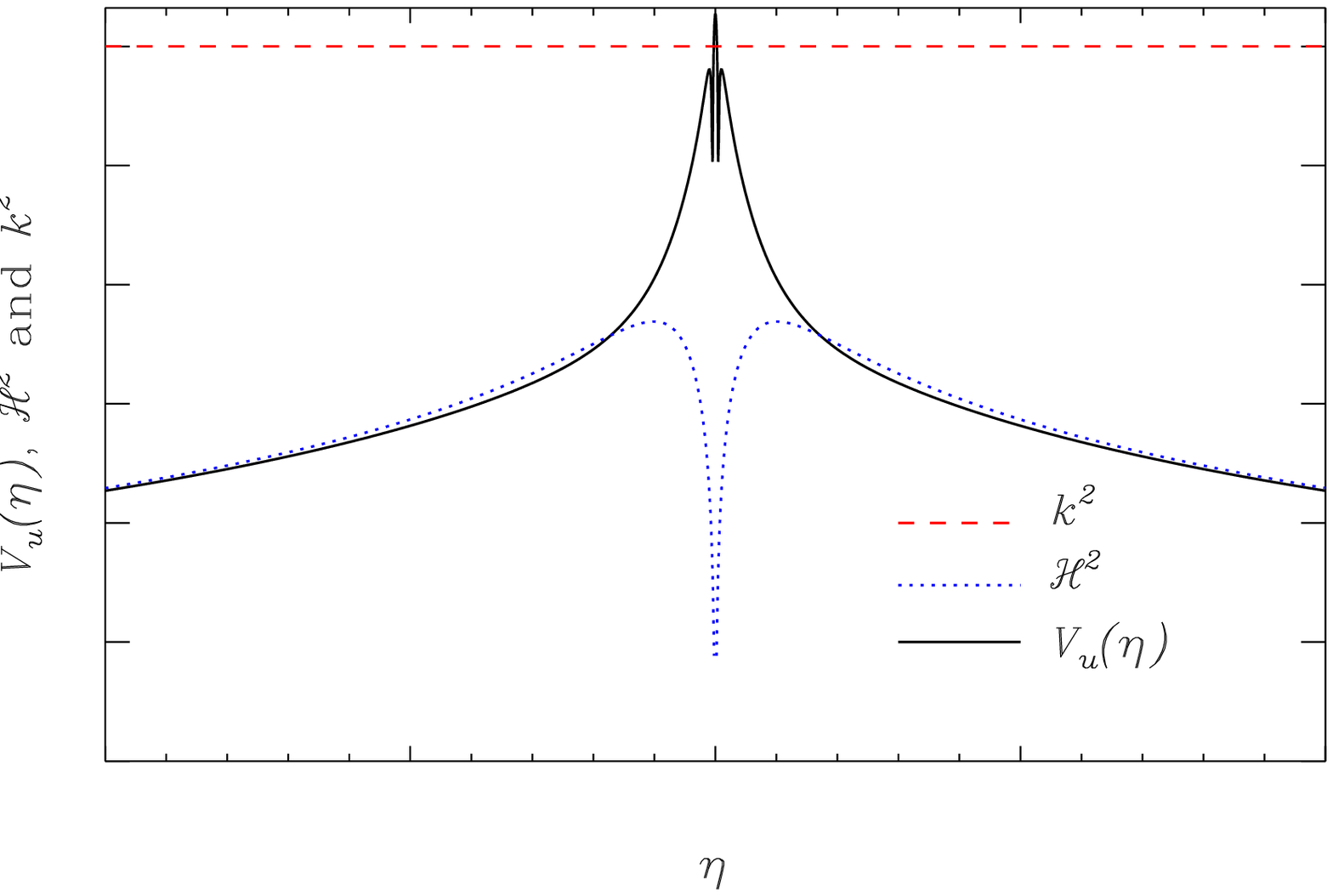}
\caption{Left panel: Effective potential $U$ and inverse horizon size
$\Hu^2$ relative to the scale $k^2$ of the perturbations in inflation
models as functions of the conformal time $\eta$. The inflation phase,
in this sketch, is smoothly linked with the radiation dominated epoch
(RDE). The times at which the effect of the potential is comparable
with the scale, \ie $k^2\sim U$ are seen to be essentially the times
at which the scale enters and exits the horizon, \ie $k\sim \Hu$, and
are hence labeled ``h.c.,'' standing for horizon crossing. The
primordial power spectrum (PPS) is understood to be the spectrum that
is obtained in the phase for which the modes are frozen and indicated
by an arrow. The actual power spectrum, in such a model, also needs to
pass the radiation to matter domination transition later on. Right
panel: Effective potential $V_u$ and inverse horizon size $\Hu^2$
relative to the scale $k^2$ of the perturbations in the bounce model
as functions of the conformal time $\eta$. The difference with the
inflation case is striking.}
\label{infRDE}
\end{center}
\end{figure*}

Another situation of cosmological interest to compare the bounce with
is a phase of quasi-exponential inflation followed by preheating and
the subsequent epoch of radiation domination. When only one field is
present, the potentials for either $u$ or $v$ are essentially
undistinguishable and both coincide numerically with the inverse
Hubble size $\Hu^2$, as shown schematically in Fig.~\ref{infRDE}. For
more than one field, the situation is qualitatively different and
cannot be understood by means of a simple potential~\cite{FB}.  For a
given wave number $k$, the spectrum is frozen when the wavelength hits
the potential, which is often phrased, because of the similarity with
the Hubble scale, as ``horizon exit'' (see Ref.~\cite{BF} for a more
detailed discussion of this point, and Ref.~\cite{MP} in the bounce
context). As illustrated in Fig.~\ref{infRDE}, the crucial difference
between the two situations, namely preheating and bounce transitions,
is that in the latter case the potential and the Hubble scale behave
in completely different ways whereas they correspond in the former, at
least in the region of potential crossing. Far from the bounce itself,
however, the potential tends to $\Hu^2$ again, in a fashion similar to
what happens during inflation. We conclude that in the bounce case,
the potential is the quantity that matters and the Hubble scale is
irrelevant for the calculation of the amplification of
perturbations. As a consequence, for practical calculatory purposes,
the phrase ``Hubble crossing'' appears misleading in this
context~\cite{MP} and the phrase ``potential crossing'' should be used
instead.

\section{Conclusions}

In this section, we summarize the main results obtained above and
discuss them in a more general framework.

Assuming general relativity as the theory describing gravitation
during a bouncing stage happening in the early universe, letting the
matter content be in the form of a scalar field, and restricting
attention to the closed spatial section case in order to satisfy the
null energy condition, we were able to develop a general formalism by
expanding any bouncing scale factor around the $\Ka=1$ de Sitter-like
bouncing solution. This expansion is characterized by two parameters
$\delta$ and $\xi$ which, in some sense, are the counterparts of the
slow-roll parameters in the usual inflationary
models~\cite{slowroll}. Because this expansion permits a general
calculation of the potential for the primordial scalar gravitational
perturbations, this allows to fully determine the structure of their
evolution as they propagate across the bounce.

The potential $V_u$ obtained is radically different from the Hubble
scale at the relevant times. This has to be contrasted with the
inflationary paradigm for which $\Hu^2$ and $V_u$ are almost
identical.

\begin{figure}
\begin{center}
\includegraphics[width=8.5cm]{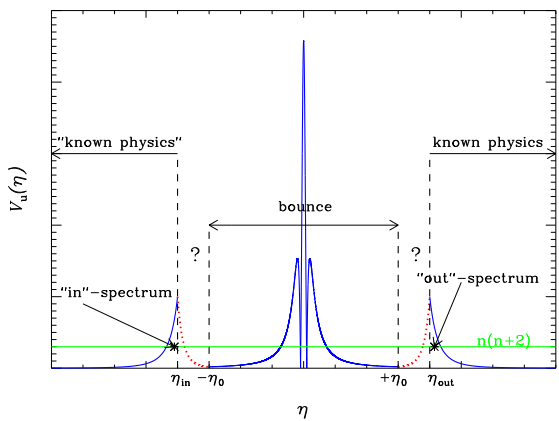}
\caption{The effective potential $V_u(\eta)$ for the perturbation
variables $u(\eta)$ for our bounce model when one connects this
bounce transition to both a previous contracting phase on one side and
to the usual radiation dominated phase later on the other side.}
\label{sketch}
\end{center}
\end{figure}

An important conclusion of this work is that a bounce phase, even a
short one, can affect large scales of perturbations. General arguments
aiming at showing the contrary therefore suffer from our
counter-example.  The bounce itself is part of the mechanism described
in the Introduction, so that the transfer matrix we obtained
participates to the one of Eq.~(\ref{transT}) through
\begin{equation}
\lim_{\eta_0\to 1}\mathbf{T} \propto \mathbf{T}^<_\mathrm{?} \cdot
k^{-1} \cdot \mathbf{T}^>_\mathrm{?}\, ,
\end{equation}
where the $k$ dependence stems from the solution (\ref{Tudelta}) and
the unknown matrices $\mathbf{T}^<_\mathrm{?}$ and
$\mathbf{T}^>_\mathrm{?}$ refer to the unknown parts sketched in
Fig.~\ref{sketch}. The coefficients one is interested in, namely
$T_{11}$ and $T_{12}$, giving the amplitude of the growing mode in the
expanding phase as functions of the modes in the contracting phase,
accordingly can depend on $k$. In addition, it is important to notice
that, as shown in Ref.~\cite{MP}, this mechanism does not violate
causality; a similar statement was also emphasized in
Ref.~\cite{weinberg}.

Paradoxically, obtaining a spectral modification at the bounce is
possible provided the bounce lasts the minimal amount of conformal
time compatible with the NEC preservation. Nevertheless, the
assumption of no effect can be justified provided the constraint
$\eta_0 -1 \not\ll 1$ is satisfied, or in the pure de Sitter case
having $\eta_0=1$ strictly. This last situation is what happens in
models in which the bounce takes place for a vanishing value of the
scalar field kinetic energy~\cite{GT}, whereas the former case implies
a kinetic energy density (not the scalar field itself) for the scalar
field comparable to the Planck scale, which may render the
semi-classical field theory dubious.

This can be particularly important in view of the string motivated
potential alternatives to inflation of the pre big bang kind if it
turns out that these models might lead to such spectral corrections as
discussed above. This condition needs be verified in each particular
situation. For instance, in the pre big bang case, one would need to
model the bounce occurring in the Einstein frame, in which our
formalism is well suited, to see what the behaviour of $V_u$ is in
this context. Therefore, and unfortunately, one consequence of the
failure of any general argument preventing any alteration of the
spectrum is that one needs to explicitly model a regime in which
higher order string corrections are dominant. Avoiding this was the
main interest of the general argument in question.

We also obtained that the relevant propagation variable is not $v$,
whose flat space equivalent is commonly used for quantization, \ie for
setting up the initial conditions, but rather the intermediate
variable $u$, directly related to the Bardeen potential. This is to be
compared with what was recently obtained in Ref.~\cite{GGV}, based on
a completely different theory of gravity, in which neither variable
happens to be bounded at the bounce.

The spectrum of gravitational wave cannot be affected by propagating
through these bounces. This exemplifies the fact that there is no
fundamental reason according to which scalar and tensor modes should
propagate similarly through a bounce.

The picture that emerges for the construction of a complete model of
the universe is shown in Fig.~\ref{sketch} and consists in a regime in
which quantum field theory in a time-dependent background is well
suited, as is the case for instance in many string motivated
scenarios~\cite{PBB,ekp}; this first phase allows an easy calculation
of a spectrum of perturbation that would be sort of
pre-primordial. Then, unless the curvature was always important in
this first period, it is followed by an unknown epoch which connects
to the bounce itself, which should also be followed by yet another
unknown epoch in order for the curvature to be
negligible~\cite{MP}. This reveals the most important difference
between bouncing scenarios and inflation, namely the need for a
high curvature phase, which we have seen may drastically modify the
physical predictions.

\vspace{3.5cm} 
\centerline{\bf Acknowledgments}
\vspace{0.2cm}

We wish to thank R.~H.~Brandenberger, A.~Buonanno, C.~Cartier,
N.~Pinto-Neto, N.~Turok, J.-P.~Uzan and F.~Vernizzi for helpful
comments. We are especially indebted to F.~Finelli for many
enlightening discussions.

\end{document}